\documentclass[acmtog,authorversion]{acmart}
\acmSubmissionID{104}

\usepackage{booktabs} 

\usepackage[ruled]{algorithm2e} 

\SetAlFnt{\small}
\SetAlCapFnt{\small}
\SetAlCapNameFnt{\small}
\SetAlCapHSkip{0pt}

\usepackage{subfigure}
\usepackage{caption}
\usepackage{algorithmic}

\usepackage{amsmath,amsfonts,bm}

\def\eqref#1{equation~\ref{#1}}

\def\1{\bm{1}}

\def\rva{{\mathbf{a}}}

\def\rvd{{\mathbf{d}}}

\def\rvg{{\mathbf{g}}}

\def\rvm{{\mathbf{m}}}

\def\rvq{{\mathbf{q}}}

\def\rvs{{\mathbf{s}}}

\def\rvv{{\mathbf{v}}}

\def\rvx{{\mathbf{x}}}

\def\rvz{{\mathbf{z}}}
\def\rvmu{{\mathbf{\mu}}}

\DeclareMathAlphabet{\mathsfit}{\encodingdefault}{\sfdefault}{m}{sl}
\SetMathAlphabet{\mathsfit}{bold}{\encodingdefault}{\sfdefault}{bx}{n}

\newcommand{\expec}{\mathbb{E}}

\acmJournal{TOG}

\setcopyright{acmlicensed}
\acmJournal{TOG}
\acmYear{2021}
\acmVolume{40}
\acmNumber{4}
\acmArticle{1}
\acmMonth{8}
\acmDOI{10.1145/3450626.3459670}

\begin{document}
\title{AMP: Adversarial Motion Priors for Stylized Physics-Based Character Control}

\author{Xue Bin Peng$^\dagger$}
\affiliation{\institution{University of California, Berkeley}}
\author{Ze Ma$^\dagger$}
\affiliation{\institution{Shanghai Jiao Tong University}}
\author{Pieter Abbeel}
\affiliation{\institution{University of California, Berkeley}}
\author{Sergey Levine}
\affiliation{\institution{University of California, Berkeley}}
\author{Angjoo Kanazawa}
\affiliation{\institution{University of California, Berkeley}}



\begin{teaserfigure}
	\centering
    \subfigure{\includegraphics[height=0.1213\textwidth]{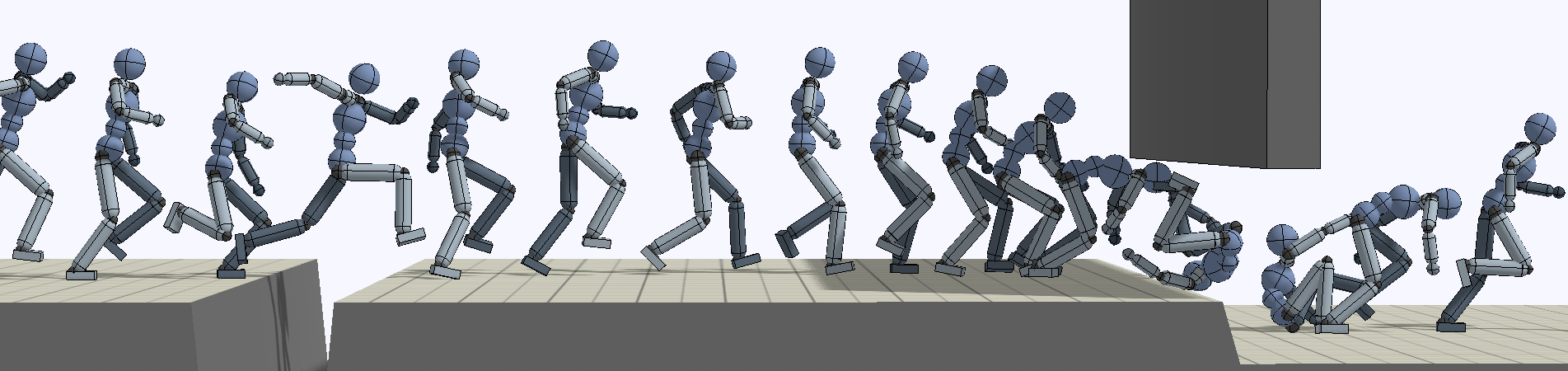}}
    \subfigure{\includegraphics[height=0.1213\textwidth]{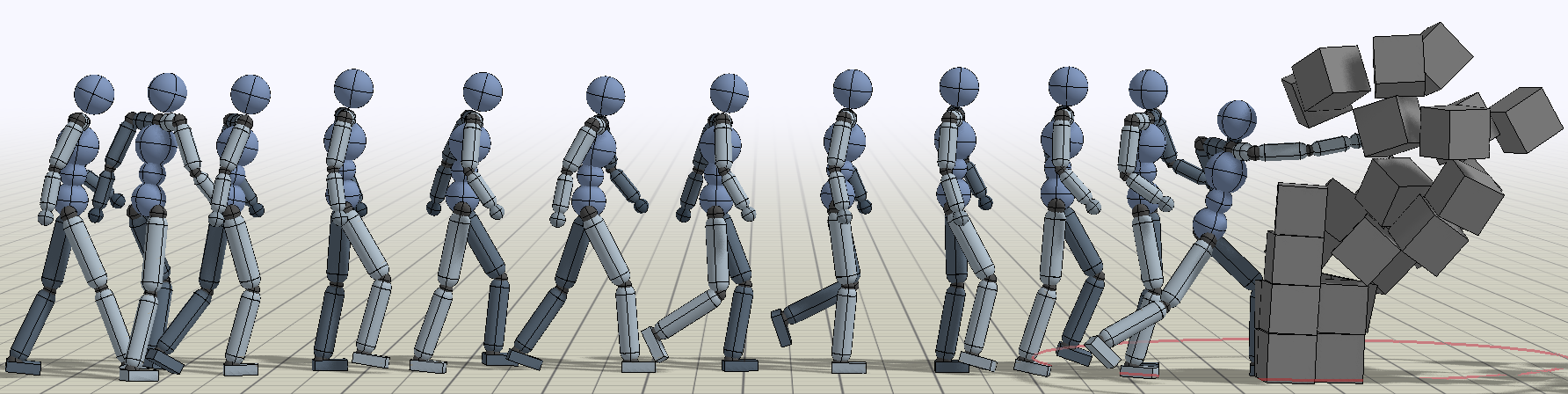}} \\
\vspace{-0.25cm}
\caption{Our framework enables physically simulated character to solve challenging tasks while adopting stylistic behaviors specified by unstructured motion data. \textbf{Left:} A character learns to traverse an obstacles course using a variety of locomotion skills. \textbf{Right:} A character learns to walk to and punch a target.}
\label{fig:teaser}
\end{teaserfigure}

\begin{abstract}
Synthesizing graceful and life-like behaviors for physically simulated characters has been a fundamental challenge in computer animation. Data-driven methods that leverage motion tracking are a prominent class of techniques for producing high fidelity motions for a wide range of behaviors. However, the effectiveness of these tracking-based methods often hinges on carefully designed objective functions, and when applied to large and diverse motion datasets, these methods require significant additional machinery to select the appropriate motion for the character to track in a given scenario.
In this work, we propose to obviate the need to manually design imitation objectives and mechanisms for motion selection by utilizing a fully automated approach based on adversarial imitation learning. High-level task objectives that the character should perform can be specified by relatively simple reward functions, while the low-level style of the character's behaviors can be specified by a dataset of unstructured motion clips, without any explicit clip selection or sequencing. For example, a character traversing an obstacle course might utilize a task-reward that only considers forward progress, while the dataset contains clips of relevant behaviors such as running, jumping, and rolling. These motion clips are used to train an adversarial motion prior, which specifies style-rewards for training the character through reinforcement learning (RL). The adversarial RL procedure automatically selects which motion to perform, dynamically interpolating and generalizing from the dataset.
Our system produces high-quality motions that are comparable to those achieved by state-of-the-art tracking-based techniques, while also being able to easily accommodate large datasets of unstructured motion clips. Composition of disparate skills emerges automatically from the motion prior, without requiring a high-level motion planner or other task-specific annotations of the motion clips. We demonstrate the effectiveness of our framework on a diverse cast of complex simulated characters and a challenging suite of motor control tasks. 

\end{abstract}

%
%


%
%


\maketitle

\let\thefootnote\relax\footnotetext{$^\dagger$ Joint first authors.}

\section{Introduction}
Synthesizing natural and life-like motions for virtual characters is a crucial element for breathing life into immersive experiences, such as films and games. The demand for realistic motions becomes even more apparent for VR applications, where users are provided with rich modalities through which to interact with virtual agents. Developing control strategies that are able to replicate the properties of naturalistic behaviors is also of interest for robotic systems, as natural motions implicitly encode important properties, such as safety and energy efficiency, which are vital for effective operation of robots in the real world. While examples of natural motions are commonplace, identifying the underlying characteristics that constitute these behaviors is nonetheless challenging, and more difficult still to replicate in a controller.

So what are the characteristics that constitute natural and life-like behaviors? Devising quantitative metrics of the \emph{naturalness} of motions has been a fundamental challenge for optimization-based character animation techniques \citep{BipedWang2009,AlBorno2013,Wampler2014}.
Heuristics such as symmetry, stability, and effort minimization can improve the realism of motions produced by physically simulated characters \citep{StyleIKGrochow2004,CIO2012,MuscleMordatch2013,SymYu2018}. But these strategies may not be broadly applicable to all behaviors of interest. Effective applications of these heuristics often require careful balancing of the various objectives, a tuning process that may need to be repeated for each task. Data-driven methods are able to mitigate some of these challenges by leveraging motion clips recorded from real-life actors to guide the behaviors of simulated characters \citep{BipedDataSok2007,DaSilva2008,Muico2009,Liu2010Samcon}. A common instantiation of this approach is to utilize a tracking objective that encourages a character to follow particular reference trajectories relevant for a given task. These tracking-based methods can produce high-quality motions for a large repertoire skills. But extending these techniques to effectively leverage large unstructured motion datasets remains challenging, since a suitable motion clip needs to be selected for the character to track at each time step. This selection process is typically performed by a motion planner, which generates reference trajectories for solving a particular task \citep{2017-TOG-deepLoco,DreCon2019,PredictSimPark2019}. However, constructing an effective motion planner can itself be a challenging endeavour, and entails significant overhead to annotate and organize the motion clips for a desired task.
For many applications, it is not imperative to exactly track a particular reference motion. Since a dataset typically provides only a limited collection of example motions, a character will inevitably need to deviate from the reference motions in order to effectively perform a given task. Therefore, the intent is often not for the character to closely track a particular motion, but to adopt general behavioral characteristics depicted in the dataset. We refer to these behavioral characteristics as a \emph{style}.

In this work, we aim to develop a system where users can specify high-level task objectives for a character to perform, while the low-level \emph{style} of a character's movements can be controlled through examples provided in the form of unstructured motion clips.
To control the style of a character's motions, we propose adversarial motion priors (AMP), a method for imitating behaviors from raw motion clips without requiring any task-specific annotations or organization of the dataset. Given a set of reference motions that constitutes a desired motion style, the motion prior is modeled as an adversarial discriminator, trained to differentiate between behaviors depicted in the dataset from those produced by the character.
The motion prior therefore acts as a general measure of similarity between the motions produced by a character and the motions in the dataset. By incorporating the motion prior in a goal-conditioned reinforcement learning framework, our system is able to train physically simulated characters to perform challenging tasks with natural and life-like behaviors. Composition of diverse behaviors emerges automatically from the motion prior, without the need for a motion planner or other mechanism for selecting \emph{which} clip to imitate.

The central contribution of this work is an adversarial learning approach for physics-based character animation that combines goal-conditioned reinforcement with an adversarial motion prior, which enables the \emph{style} of a character's movements to be controlled via example motion clips, while the \emph{task} is specified through a simple reward function. We present one of the first adversarial learning systems that is able to produce high-quality full-body motions for physically simulated characters.
By combining the motion prior with additional task objectives, our system provides a convenient interface through which users can specify high-level directions for controlling a character's behaviors. These task objectives allow our characters to acquire more complex skills than those demonstrated in the original motion clips.
While our system is built on well-known adversarial imitation learning techniques, we propose a number of important design decisions that lead to substantially higher quality results than those achieved by prior work, enabling our characters to learn highly dynamic and diverse motors skills from unstructured motion data.

\section{Related Work}

Developing systems that can synthesize natural motions for virtual characters is one of the fundamental challenges of computer animation. These procedural animation techniques can be broadly categorized as \emph{kinematic methods} and \emph{physics-based methods}. Kinematic methods generally do not explicitly utilize the equations of motion for motion synthesis. Instead, these methods often leverage datasets of motion clips to generate motions for a character \citep{MotionLee2002,MotionFields2010}. Given a motion dataset, controllers can be constructed to select an appropriate motion clip to play back for a particular scenario \citep{MotionGraph2007,Treuille2007,2016-TOG-taskBasedLocomotion}. Data-driven methods using generative models, such as Gaussian processes \citep{Ye2010,2012-ccclde} and neural networks \citep{2020-TOG-MVAE,PFNN2017,MANN2018}, have also been applied to synthesize motions online. When provided with sufficiently large and high-quality datasets, kinematic methods are able to produce realistic motions for a large variety of sophisticated skills \citep{MotionFields2010,Levine2011,2016-TOG-taskBasedLocomotion,Lee2018,NSM2019}. However, their ability to synthesize motions for novel situations can be limited by the availability of data. For complex tasks and environments, it can be difficult to collect a sufficient amount of data to cover all possible behaviors that a character may need to perform.
This is particularly challenging for nonhuman and fictional creatures, where motion data can be scarce. In this work, we combine data-driven techniques with physics-based animation methods to develop characters that produce realistic and responsive behaviors to novel tasks and environments.

\paragraph{Physics-Based Methods:} Physics-based methods address some of the limitations of kinematic methods by synthesizing motions from first principles. These methods typically leverage a physics simulation, or more general knowledge of the equations of motion, to generate motions for a character \citep{Raibert1991,Wampler2014}. Optimization techniques, such as trajectory optimization and reinforcement learning, play a pivotal role in many physics-based methods, where controllers that drive a character's motions are produced by optimizing an objective function \citep{Panne94virtualwindup,CIO2012,BicycleTan2014}. While these methods are able to synthesize physically plausible motions for novel scenarios, even in the absence of motion data, designing effective objectives that lead to natural behaviors can be exceptionally difficult. Heuristics derived from prior knowledge of the characteristics of natural motions are commonly included into the objective function, such as symmetry, stability, effort minimization, and many more \citep{BipedWang2009,CIO2012,SymYu2018}. Simulating more biologically accurate actuators can also improve motion quality \citep{MuscleWang2012,2013-TOG-MuscleBasedBipeds,MuscleJiang2019}, but may nonetheless yield unnatural behaviors.

\paragraph{Imitation Learning:}
The challenges of designing objective functions that lead to natural motions have spurred the adoption of data-driven physics-based animation techniques \citep{Zordan2002,Sharon2005Walking,DaSilva2008,DataDrivenLee2010,PendulumKwon2017}, which utilizes reference motion data to improve motion quality. Reference motions are typically incorporated through an imitation objective that encourages a character to imitate motions in the dataset. The imitation objective is commonly implemented as a tracking objective, which attempts to minimize the pose error between the simulated character and target poses from a reference motion \citep{BipedDataSok2007,DataDrivenLee2010,Liu2010Samcon,2016-TOG-controlGraphs,2018-TOG-deepMimic}. Since the pose error is generally computed with respect to a single target pose at a time, some care is required to select an appropriate target pose from the dataset.
A simple strategy is to synchronize the simulated character with a given reference motion using a phase variable \citep{2018-TOG-deepMimic,2018-TOG-SFV,MuscleLee2019}, which is provided as an additional input to the controller. The target pose at each time step can then be conveniently determined by selecting the target pose according to the phase. This strategy has been effective for imitating individual motion clips, but it can be difficult to scale to datasets containing multiple disparate motions, as it may not be possible to synchronize and align multiple reference motions according to a single-phase variable. Recent methods have extended these tracking-based techniques to larger motion datasets by explicitly providing target poses from the reference motion that is being tracked as inputs to the controller \citep{MocapImitationChentanez2018,DreCon2019,PredictSimPark2019,Won2020}. This then allows a controller to imitate different motions depending on the input target poses. However, selecting the appropriate motion for a character to imitate in a given scenario can still entail significant algorithmic overhead. These methods often require a high-level motion planner that selects which motion clip the character should imitate for a given task \citep{2017-TOG-deepLoco,DreCon2019,PredictSimPark2019}. The character's performance on a particular task can therefore be constrained by the performance of the motion planner.

Another major limitation of tracking-based imitation techniques is the need for a pose error metric when computing the tracking objective \citep{Sharon2005Walking,Liu2010Samcon,2018-TOG-deepMimic}. 
These error metrics are often manually-designed, and it can be challenging to construct and tune a common metric that is effective across all skills that a character is to imitate. Adversarial imitation learning provides an appealing alternative \citep{ApprenticeAbbeel2004,MaxEntIRL2008,GAIL2016}, where instead of using a manually-designed imitation objective, these algorithms train an adversarial discriminator to differentiate between behaviors generated by an agent from behaviors depicted in the demonstration data (e.g. reference motions). The discriminator then serves as the objective function for training a control policy to imitate the demonstrations. While these methods have shown promising results for motion imitation tasks \citep{Merel2017,DiverseImitationWang2017}, adversarial learning algorithms can be notoriously unstable and the resulting motion quality still falls well behind what has been achieved with state-of-the-art tracking-based techniques. 
\citet{peng2018variational} was able to able to produce substantially more realistic motions by regularizing the discriminator with an information bottleneck. However, their method still requires a phase variable to synchronize the policy and discriminator with the reference motion. Therefore, their results are limited to imitating a single motion per policy, and thus not suitable for learning from large diverse motion datasets. In this work, we propose an adversarial method for learning general motion priors from large unstructured datasets that contain diverse motion clips. Our approach does not necessitate any synchronization between the policy and reference motion. Furthermore, our approach does not require a motion planner, or any task-specific annotation and segmentation of the motion clips \citep{2017-TOG-deepLoco,PredictSimPark2019,DreCon2019}. Instead, composition of multiple motions in furtherance of a task objective emerges automatically through the motion prior. We also present a number of design decisions for stabilizing the adversarial training process, leading to consistent and high-quality results.

\paragraph{Latent Space Models:}
Latent space models can also act as a form of motion prior that leads to more life-like behaviors. These models specify controls through a learned latent representation, which is then mapped to controls for the underlying system \citep{OmniBurgard2008,HeessWTLRS16,ICLR17-Florensa,hausman2018learning}. The latent representation is typically learned through a pre-training phase using supervised learning or reinforcement learning techniques to encode a diverse range of behaviors into a latent representation. Once trained, this latent representation can be used to build a control hierarchy, where the latent space model acts as a low-level controller, and a separate high-level controller is trained to specify controls via the latent space \citep{ICLR17-Florensa,haarnoja18a,lynch20a}. For motion control of simulated characters, the latent representation can be trained to encode behaviors from reference motion clips, which then constrains the behavior of a character to be similar to those observed in the motion data, therefore leading to more natural behaviors for downstream tasks \citep{merel2019neural,MCPPeng19}. However, since the realism of the character's motions is enforced implicitly through the latent representation, rather than explicitly through an objective function, it is still possible for the high-level controller to specify latent encodings that produce unnatural behaviors \citep{MCPPeng19,CatchCarryMerel2020}. \citet{CARLLuo2020} proposed an adversarial domain confusion loss to prevent the high-level controller from specifying encodings that are different from those observed during pre-training. However, since this adversarial objective is applied in the latent space, rather than on the actual motions produced by the character, the model is nonetheless prone to generating unnatural behaviors. Our proposed motion prior directly enforces similarity between the motions produced by the character and those in the reference motion dataset, which enables our method to produce higher fidelity motions than what has been demonstrated by latent space models. Our motion prior also does not require a separate pre-training phase, and instead, can be trained jointly with the policy.

\section{Overview}
Given a dataset of reference motions and a task objective defined by a reward function, our system synthesizes a control policy that enables a character to achieve the task objective in a physically simulated environment, while utilizing behaviors that resemble the motions in the dataset. Crucially, the character's behaviors need not exactly match any specific motion in the dataset, instead its movements need only to adopt more general characteristics exhibited by the corpus of reference motions. These reference motions collectively provide an example-based definition of a behavioral \emph{style}, and by providing the system with different motion datasets, the character can then be trained to perform a task in a variety of distinct styles.

\begin{figure}[t]
	\centering
    \includegraphics[width=1\columnwidth]{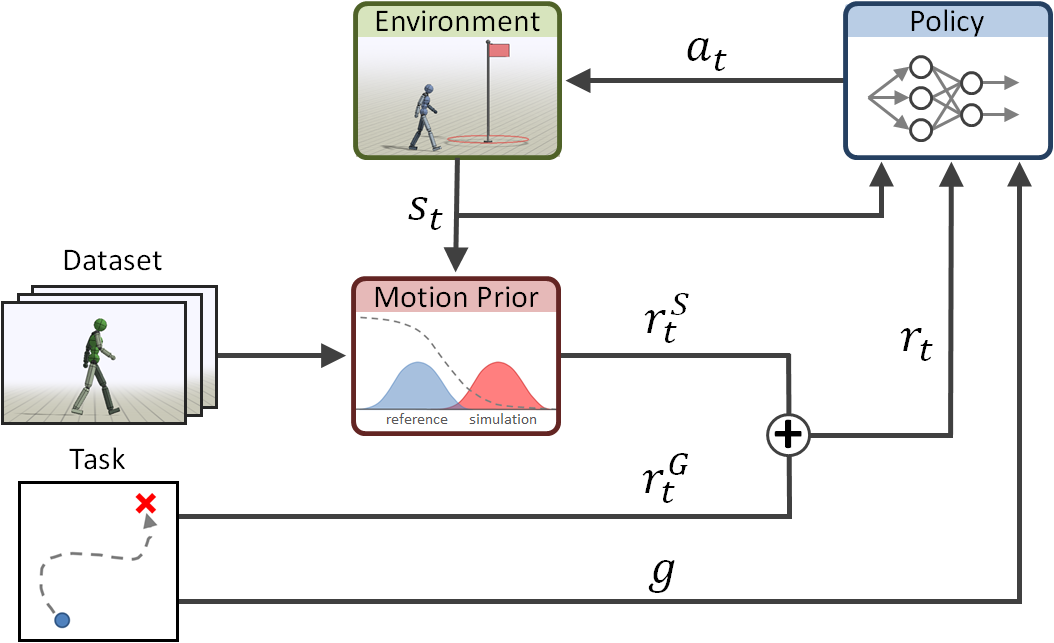}\\
    \vspace{-0.2cm}
\caption{Schematic overview of the system. Given a motion dataset defining a desired motion style for the character, the system trains a motion prior that specifies style-rewards $r_t^S$ for the policy during training. These style-rewards are combined with task-rewards $r_t^G$ and used to train a policy that enables a simulated character to satisfy task-specific goals $\rvg$, while also adopting behaviors that resemble the reference motions in the dataset.}
\label{fig:overview}
\vspace{-0.3cm}
\end{figure}

Figure~\ref{fig:overview}
provides a schematic overview of the system. The motion dataset $\mathcal{M}$ consists of a collection of reference motions, where each motion $\rvm^i = \{\hat{\rvq}_t^i\}$ is represented as a sequence of poses $\hat{\rvq}_t^i$. The motion clips may be collected from the mocap of real-life actors or from artist-authored keyframe animations. Unlike previous frameworks, our system can be applied directly on raw motion data, without requiring task-specific annotations or segmentation of a clip into individual skills.
The motion of the simulated character is controlled by a policy $\pi(\rva_t | \rvs_t, \rvg)$ that maps the state of the character $\rvs_t$ and a given goal $\rvg$ to a distribution over actions $\rva_t$. The actions from the policy specify target positions for proportional-derivative (PD) controllers positioned at each of the character's joints, which in turn produce control forces that drive the motion of the character. The goal $\rvg$ specifies a task reward function $r^G_t = r^G(\rvs_t, \rva_t, \rvs_{t+1}, \rvg)$, which defines high-level objectives for the character to satisfy (e.g. walking in a target direction or punching a target). The style objective $r^S_t = r^S(\rvs_t, \rvs_{t+1})$ is specified by an adversarial discriminator, trained to differentiate between motions depicted in the dataset from motions produced by the character. The style objective therefore acts as a task-agnostic motion prior that provides an a-priori estimate of the naturalness or style of a given motion, independent of a specific task. The style objective then encourages the policy to produce motions that resemble behaviors depicted in the dataset.

\vspace{-0.05cm}
\section{Background}
Our system combines techniques from goal-conditioned reinforcement learning and generative adversarial imitation learning to train control policies that enable simulated characters to perform challenging tasks in a desired behavioral style. In this section, we provide a brief review of these techniques.

\vspace{-0.1cm}
\subsection{Goal-Conditioned Reinforcement Learning}
Our characters are trained through a goal-conditioned reinforcement learning framework, where an agent interacts with an environment according to a policy $\pi$ in order to fulfill a given goal $\rvg \in \mathcal{G}$ sampled according to a goal distribution ${\rvg \sim p(\rvg)}$. At each time step $t$, the agent observes the state $\rvs_t \in \mathcal{S}$ of the system, then samples an action $\rva_t \in \mathcal{A}$ from a policy $\rva_t \sim \pi(\rva_t | \rvs_t, \rvg)$. The agent then applies that action, which results in a new state $\rvs_{t+1}$, as well as a scalar reward $r_t = r(\rvs_t, \rva_t, \rvs_{t+1}, \rvg)$. The agent's objective is to learn a policy that maximizes its expected discounted return $J(\pi)$,
\begin{equation}
    J(\pi) = \expec_{p(\rvg)}\expec_{p(\tau | \pi, \rvg)} \left[ \sum_{t=0}^{T-1} \gamma^t r_t \right],
\label{eqn:rl_objective}
\end{equation}
where $p(\tau | \pi, \rvg) = p(\rvs_0) \prod_{t=0}^{T-1} p(\rvs_{t+1} | \rvs_t, \rva_t) \pi(\rva_t | \rvs_t, \rvg)$ represents the likelihood of a trajectory $\tau = \{(\rvs_t, \rva_t, r_t)_{t=0}^{T-1}, \rvs_T \}$ under a policy $\pi$ for a goal $\rvg$. $p(\rvs_0)$ is the initial state distribution, and $p(\rvs_{t+1} | \rvs_t, \rva_t)$ represents the dynamics of the environment. $T$ denotes the time horizon of a trajectory, and $\gamma \in [0, 1)$ is a discount factor.

\subsection{Generative Adversarial Imitation Learning}
Generative adversarial imitation learning (GAIL) \citep{GAIL2016} adapts techniques developed for generative adversarial networks (GAN) \citep{GAN2014} to the domain of imitation learning. In the interest of brevity, we exclude the goal $\rvg$ from the notation, but the following discussion readily generalizes to goal-conditioned settings. Given a dataset of demonstrations $\mathcal{M} = \{(\rvs_i, \rva_i)\}$, containing states $\rvs_i$ and actions $\rva_i$ recorded from an unknown demonstration policy, the objective is to train a policy $\pi(\rva | \rvs)$ that imitates the behaviors of the demonstrator. Behavioral cloning can be used to directly fit a policy to map from states observed in $\mathcal{M}$ to their corresponding actions using supervised learning \citep{Pomerleau1998,Bojarski2016}. However, if only a small amount of demonstrations are available, then behavioral cloning techniques are prone to drift \citep{DaggerRoss2011}. Furthermore, behavioral cloning is not directly applicable in settings where the demonstration actions are not observable (e.g. reference motion data).

GAIL addresses some of the limitations of behavioral cloning by learning an objective function that measures the similarity between the policy and the demonstrations, and then updating $\pi$ via reinforcement learning to optimize the learned objective. The objective is modeled as a discriminator $D(\rvs, \rva)$, trained to predict whether a given state $\rvs$ and action $\rva$ is sampled from the demonstrations $\mathcal{M}$ or generated by running the policy $\pi$,
\begin{align}
    \mathop{\mathrm{arg \ min}}_D \ -\expec_{d^\mathcal{M}(\rvs, \rva)} \left[ \mathrm{log}\left(D(\rvs, \rva)\right) \right] - \expec_{d^\pi({\rvs, \rva})} \left[\mathrm{log}\left(1 - D(\rvs, \rva)\right) \right].
\label{eqn:disc_loss}
\end{align}
$d^\mathcal{M}(\rvs, \rva)$ and $d^\pi({\rvs, \rva})$ denote the state-action distribution of the dataset and policy respectively. The policy is then trained using the RL objective detailed in Equation~\ref{eqn:rl_objective}, with rewards specified by,
\begin{equation}
    r_t = -\mathrm{log}\left(1 - D(\rvs_t, \rva_t)\right).
\end{equation}
This adversarial training procedure can be interpreted as training a policy to produce states and actions that appear to the discriminator as being indistinguishable from the demonstrations. It can be shown that this objective minimizes the Jensen-Shannon divergence between $d^\mathcal{M}(\rvs, \rva)$ and $d^\pi({\rvs, \rva})$ \citep{FGAN2016,FDivKe2019}.

\section{Adversarial Motion Prior}
In this work, we consider reward functions that consist of two components specifying: 1) \emph{what} task a character should perform, and 2) \emph{how} the character should go about performing that task,
\begin{equation}
    r(\rvs_t, \rva_t, \rvs_{t+1}, \rvg) = w^G r^G(\rvs_t, \rva_t, \rvs_t, \rvg) + w^S r^S(\rvs_t, \rvs_{t+1}).
\label{eqn:reward}
\end{equation}
The \emph{what} is represented by a task-specific reward $r^G(\rvs_t, \rva_t, \rvs_t, \rvg)$, which defines high-level objectives that a character should satisfy (e.g. moving to a target location). The \emph{how} is represented through a learned task-agnostic style-reward $r^S(\rvs_t, \rvs_{t+1})$, which specifies low-level details of the behaviors that the character should adopt when performing the task (e.g., walking vs. running to a target). The two reward terms are combined linearly with weights $w^G$ and $w^S$. The task-reward $r^G$ can be relatively intuitive and simple to design. However, it can be exceptionally difficult to design a style-reward $r^S$ that leads a character to learn naturalistic behaviors, or behaviors that conform to a particular style. Learning effective style objectives will therefore be the primary focus of this work.

We propose to model the style-reward with a learned discriminator, which we refer to as an adversarial motion prior (AMP), by analogy to the adversarial pose priors that were previously proposed for vision-based pose estimation tasks \citep{hmrKanazawa17}.
Unlike standard tracking objectives, which measure pose similarity with respect to a specific reference motion, the motion prior returns a general score indicating the similarity of the character's motion to the motions depicted in the dataset, without explicitly comparing to a particular motion clip. Given a motion dataset, the motion prior is trained using the GAIL framework to predict whether a state transition $(\rvs_t, \rvs_{t+1})$ is a \emph{real} sample from the dataset or a \emph{fake} sample produced by the character. The motion prior is independent of the task-specific goal $\rvg$, therefore a single motion prior can be applied to multiple tasks, and different motion priors can be applied to train policies that perform the same task but in different styles. By combining GAIL with additional task objectives, our approach decouples task specification from style specification, thereby enabling our characters to perform tasks that may not be depicted in the original demonstrations.
However, adversarial RL techniques are known to be highly unstable. In the following sections, we discuss a number of design decisions to stabilize the training process and produce higher fidelity results.

\subsection{Imitation from Observations}
The original formulation of GAIL requires access to the demonstrator's actions \citep{GAIL2016}. However, when the demonstrations are provided in the form of motion clips, the actions taken by the demonstrator are unknown, and only states are observed in the data. To extend GAIL to settings with state-only demonstrations, the discriminator can be trained on state transitions $D(\rvs, \rvs')$ instead of state-action pairs $D(\rvs, \rva)$ \citep{GAILfOFaraz2018},
\begin{align}
    \mathop{\mathrm{arg \ min}}_D & -\expec_{ d^\mathcal{M}(\rvs, \rvs')} \left[ \mathrm{log}\left(D(\rvs, \rvs')\right) \right]
    - \expec_{d^\pi({\rvs, \rvs'})} \left[\mathrm{log}\left(1 - D(\rvs, \rvs')\right) \right].
\label{eqn:disc_loss_trans}
\end{align}
$d^\mathcal{M}(\rvs, \rvs')$ and $d^\pi(\rvs, \rvs')$ denote the likelihood of observing a state transition from $\rvs$ to $\rvs'$ in the dataset $\mathcal{M}$ and by following policy $\pi$ respectively. Note that if the demonstrator is different from the agent (e.g. a human actor), the observed state transitions may not be physically consistent for the agent, and therefore impossible for the agent to perfectly reproduce. Despite this discrepancy, we show that the discriminator still provides an effective objective for imitating a wide range of behaviors.

\subsection{Least-Squares Discriminator}
The standard GAN objective detailed in Equation~\ref{eqn:disc_loss_trans} typically uses a sigmoid cross-entropy loss function. However, this loss tends to lead to optimization challenges due to vanishing gradients as the sigmoid function saturates, which can hamper training of the policy \citep{WGANArjovsky2017}. A myriad of techniques have been proposed to address these optimization challenges \citep{DCGAN15,SalimansGZCRC16,ProgressiveGAN17,WGANArjovsky2017,BEGAN2017,DRAGAN2017,ImpWGAN2017,mescheder18a}. In this work, we adopt the loss function proposed for least-squares GAN (LSGAN) \citep{LSGANMao2017}, which has demonstrated more stable training and higher quality results for image synthesis tasks. The following objective is used to train the discriminator,
\begin{align}
    \mathop{\mathrm{arg \ min}}_D \ \ \expec_{d^\mathcal{M}(\rvs, \rvs')} \left[ \left(D(\rvs, \rvs') - 1 \right)^2 \right] + \expec_{d^\pi(\rvs, \rvs')} \left[\left(D(\rvs, \rvs') + 1 \right)^2 \right].
\label{eqn:disc_loss_ls}
\end{align}
The discriminator is trained by solving a least-squares regression problem to predict a score of $1$ for samples from the dataset and $-1$ for samples recorded from the policy. The reward function for training the policy is then given by
\begin{equation}
    r(\rvs_t, \rvs_{t+1}) = \mathrm{max}\left[0, \ \ 1 - 0.25 (D(\rvs_t, \rvs_{t+1}) - 1)^2 \right] .
\label{eqn:ls_gan_reward}
\end{equation}
The additional offset, scaling, and clipping are applied to bound the reward between $[0, 1]$, as is common practice in previous RL frameworks \citep{2016-TOG-deepRL,2018-TOG-deepMimic,DMControl2018}. \citet{LSGANMao2017} showed that this least-squares objective minimizes the Pearson $\chi^2$ divergence between $d^\mathcal{M}(\rvs, \rvs')$ and $d^\pi(\rvs, \rvs')$.

\subsection{Discriminator Observations}

Since the discriminator specifies rewards for training the policy, selecting an appropriate set of features for use by the discriminator when making its predictions is vital to provide the policy with effective feedback. As such, before a state transition is provided as input to the discriminator, we first apply an observation map $\Phi(\rvs_t)$ that extracts a set of features relevant for determining the characteristics of a given motion. The resulting features are then used as inputs to the discriminator $D(\Phi(\rvs), \Phi(\rvs'))$.
The set of features include:
\begin{itemize}
    \item Linear velocity and angular velocity of the root, represented in the character's local coordinate frame.
    \item Local rotation of each joint.
    \item Local velocity of each joint.
    \item 3D positions of the end-effectors (e.g. hands and feet), represented in the character's local coordinate frame.
\end{itemize}
The root is designated to be the character's pelvis. The character's local coordinate frame is defined with the origin located at the root, the x-axis oriented along the root link's facing direction, and the y-axis aligned with the global up vector. The 3D rotation of each spherical joint is encoded using two 3D vectors corresponding to the normal and tangent in the coordinate frame. This rotation encoding provides a smooth and unique representation of a given rotation. This set of observation features for the discriminator is selected to provide a compact representation of the motion across a single state transition. The observations also do not include any task-specific features, thus enabling the motion prior to be trained without requiring task-specific annotation of the reference motions, and allowing motion priors trained with the same dataset to be used for different tasks.

\subsection{Gradient Penalty}
The interplay between the discriminator and generator in a GAN often results in unstable training dynamics. One source of instability is due to function approximation errors in the discriminator, where the discriminator may assign nonzero gradients on the manifold of real data samples \citep{mescheder18a}. These nonzero gradients can cause the generator to \emph{overshoot} and move off the data manifold, instead of converging to the manifold, leading to oscillations and instability during training. To mitigate this phenomenon, a gradient penalty can be applied to penalize nonzero gradients on samples from the dataset \citep{DRAGAN2017,ImpWGAN2017,mescheder18a}. We incorporate this technique to improve training stability. The discriminator objective is then given by:
\begin{align}
    \mathop{\mathrm{arg \ min}}_D \ & \quad \expec_{d^\mathcal{M}\left(\rvs, \rvs'\right)} \left[ \left(D(\Phi(\rvs), \Phi(\rvs')) - 1 \right)^2 \right] \nonumber \\ 
    & + \expec_{d^\pi({\rvs, \rvs'})} \left[\left(D\left(\Phi(\rvs), \Phi(\rvs')\right) + 1 \right)^2 \right] \nonumber \\
    & + \frac{w^\mathrm{gp}}{2} \ \expec_{d^\mathcal{M}\left(\rvs, \rvs'\right)} \left[ \left|\left|\nabla_\phi D(\phi) \middle|_{\phi = \left(\Phi(\rvs),\Phi( \rvs')\right)} \right|\right|^2 \right],
\label{eqn:disc_loss_gp}
\end{align}
where $w^\mathrm{gp}$ is a manually-specified coefficient. Note, the gradient penalty is calculated with respect to the observation features $\phi = \left(\Phi(\rvs),\Phi( \rvs')\right)$, not the full set of state features $(\rvs, \rvs')$. As we show in our experiments, the gradient penalty is crucial for stable training and effective performance.

\section{Model Representation}

Given a high-level task objective and a dataset of reference motions, the agent is responsible for learning a control policy that fulfills the task objectives, while utilizing behaviors that resemble the motions depicted in the dataset. In this section, we detail the design of various components of the learning framework.

\subsection{States and Actions}
The state $\rvs_t$ consists of a set of features that describes the configuration of the character's body. The features are similar to those used by \citet{2018-TOG-deepMimic}, which include the relative positions of each link with respective to the root, the rotation of each link as represented using the 6D normal-tangent encoding, along with the link's linear and angular velocities. All features are recorded in the character's local coordinate system. Unlike previous systems, which synchronize the policy with a particular reference motion by including additional phase information in the state, such as scalar phase variables \citep{2018-TOG-deepMimic,MuscleLee2019} or target poses \citep{MocapImitationChentanez2018,DreCon2019,Won2020}, our policies are not trained to explicitly imitate any specific motion from the dataset. Therefore, no such synchronization or phase information is necessary.

Each action $\rva_t$ specifies target positions for PD controllers positioned at each of the character's joints. For spherical joints, each target is specified in the form of a 3D exponential map $\rvq \in \mathbb{R}^3$ \citep{ExpMapGrassia1998}, where the rotation axis $\rvv$ and rotation angle $\theta$ can be determined according to:
\begin{equation}
    \rvv = \frac{\rvq}{||\rvq||_2}, \qquad \theta = ||\rvq||_2.
\end{equation}
This representation provides a more compact parameterization than the 4D axis-angle or quaternion representations used in prior systems \citep{2018-TOG-deepMimic,Won2020}, while also avoiding gimbal lock from parameterizations such as euler angles. Target rotations for revolute joints are specified as 1D rotation angles $q = \theta$.

\begin{algorithm}[t!]
\caption{Training with AMP}
\label{alg:AMP}
\begin{algorithmic}[1]
\STATE{{\bf input} $\mathcal{M}$: dataset of reference motions}

\STATE{$D \leftarrow$ initialize discriminator}
\STATE{$\pi \leftarrow$ initialize policy}
\STATE{$V \leftarrow$ initialize value function}
\STATE{$\mathcal{B} \leftarrow \emptyset$ \ initialize reply buffer}

\item[]
\WHILE{not done}
    \FOR{trajectory $i = 1,...,m$}
    	\STATE{$\tau^i \leftarrow \{(\rvs_t, \rva_t, r^G_t)_{t=0}^{T-1}, \ \rvs^G_T, \rvg \}$ collect trajectory with $\pi$}
        \FOR{time step $t = 0,...,T-1$}
            \STATE{$d_t \leftarrow D(\Phi(\rvs_t), \Phi(\rvs_{t+1}))$}
            \STATE{$r^S_t \leftarrow $ calculate style reward according to Equation~\ref{eqn:ls_gan_reward} using $d_t$}
            \STATE{$r_t \leftarrow w^G r^G_t + w^S r^S_t$}
            \STATE{record $r_t$ in $\tau^i$}
        \ENDFOR
        \STATE{store $\tau^i$ in $\mathcal{B}$}
    \ENDFOR
    
    \item[]
    \FOR{update step $= 1,...,n$}
        \STATE{$b^{\mathcal{M}} \leftarrow$ sample batch of $K$ transitions $\{(\rvs_j, \rvs'_j)\}_{j=1}^K$ from $\mathcal{M}$}
        \STATE{$b^\pi \leftarrow$ sample batch of $K$ transitions $\{(\rvs_j, \rvs'_j)\}_{j=1}^K$ from $\mathcal{B}$}
        \STATE{update $D$ according to Equation~\ref{eqn:disc_loss_gp} using $b^{\mathcal{M}}$ and $b^\pi$}
    \ENDFOR
    
    \item[]
    \STATE{update $V$ and $\pi$ using data from trajectories $\{\tau^i\}_{i=1}^m$}
        
\ENDWHILE
\end{algorithmic}
\end{algorithm}

\subsection{Network Architecture}
Each policy $\pi$ is modeled by a neural network that maps a given state $\rvs_t$ and goal $\rvg$ to a Gaussian distribution over actions $\pi(\rva_t | \rvs_t, \rvg) = \mathcal{N}\left(\rvmu(\rvs_t, \rvg), \Sigma \right)$, with an input-dependent mean $\rvmu(\rvs_t, \rvg)$ and a fixed diagonal covariance matrix $\Sigma$. The mean is specified by a fully-connected network with two hidden layers, consisting of 1024 and 512 ReLU \citep{ReLUNair2010}, followed by a linear output layer. The values of the covariance matrix $\Sigma=\mathrm{diag}(\sigma_1, \sigma_2, ...)$ are manually-specified and kept fixed over the course of training. The value function $V(\rvs_t, \rvg)$ and discriminator $D(\rvs_t, \rvs_{t+1})$ are modeled by separate networks with a similar architecture as the policy.

\subsection{Training}
Our policies are trained using a combination of GAIL \citep{GAIL2016} and proximal-policy optimization (PPO) \citep{PPO2017}. 

Algorithm~~\ref{alg:AMP} provides an overview of the training process. At each time step $t$, the agent receives a task-reward $r^G_t = r^G(\rvs_t, \rva_t, \rvs_{t+1}, \rvg)$ from the environment, it then queries the motion prior for a style-reward $r^S_t = r^S(\rvs_t, \rvs_{t+1})$, computed according to Equation~\ref{eqn:ls_gan_reward}. The two rewards are combined according to Equation~\ref{eqn:reward} to yield the reward for the particular timstep. Following the approach proposed by \citet{2018-TOG-deepMimic}, we incorporate reference state initialization and early termination. Reference state initialization is applied by initializing the character to states sampled randomly from all motion clips in the dataset. Early termination is triggered on most tasks when any part of the character's body, with exception of the feet, makes contact with the ground. This termination criteria is disabled for more contact-rich tasks, such as rolling or getting up after a fall.

Once a batch of data has been collected with the policy, the recorded trajectories are used to update the policy and value function. The value function is updated with target values computed using TD($\lambda$) \citep{Sutton1998}. The policy is updated using advantages computed using GAE($\lambda$) \citep{GAESchulman2015}. Each trajectory recorded from the policy is also stored in a replay buffer $\mathcal{B}$, containing trajectories from past training iterations. The discriminator is updated according to Equation~\ref{eqn:disc_loss_gp} using minibatches of transitions $(\rvs, \rvs')$ sampled from the reference motion data set $\mathcal{M}$ and transitions from the replay buffer $\mathcal{B}$. The replay buffer helps to stabilize training by preventing the discriminator from overfitting to the most recent batch of trajectories from the policy.

\section{Tasks}
To evaluate AMP's effectiveness for controlling the style of a character's motions, we apply our framework to train complex 3D simulated characters to perform various motion control tasks using different motion styles. The characters include a 34 DoF humanoid, a 59 DoF T-Rex, and a 64 DoF dog. A summary of each task is provided below. Please refer to Appendix~\ref{supp:tasks} for a more in-depth description of each task and their respective reward functions.

\paragraph{Target Heading:}
In this task, the character's objective is to move along a target heading direction $\rvd^*$ at a target speed $v^*$. The goal for the policy is specified as $\rvg_t = (\tilde{\rvd}^*_t, v^*)$, with $\tilde{\rvd}^*_t$ being the target direction in the character's local coordinate frame. The target speed is selected randomly between $v^* \in [1, 5]$m/s. For slower moving styles, such as Zombie and Stealthy, the target speed is fixed at $1$m/s.

\paragraph{Target Location:} In this task, the character's objective is to move to a target location $\rvx^*$. The goal $\rvg_t = \tilde{\rvx}^*_t$ records the target location in the character's local coordinate frame.

\paragraph{Dribbling:}
To evaluate our system on more complex object manipulation tasks, we train policies for a dribbling task, where the character's objective is to dribble a soccer ball to a target location. The goal $\rvg_t = \tilde{\rvx}^*_t$ records the relative position of the target location with respect to the character. The state $\rvs_t$ is augmented with additional features that describe the state of the ball, including the position $\tilde{\rvx}^\mathrm{ball}_t$, orientation $\tilde{\rvq}^\mathrm{ball}_t$, linear velocity $\tilde{\dot{\rvx}}^\mathrm{ball}_t$, and angular velocity $\tilde{\dot{\rvq}}^\mathrm{ball}_t$ of the ball in the character's local coordinate frame.

\paragraph{Strike:}
To demonstrate AMP's ability to compose diverse behaviors, we consider a task where the character's objective is to strike a target using a designated end-effector (e.g. hands). The target may be located at various distances from the character. Therefore, the character must first move close to the target before striking it. These distinct phases entail different optimal behaviors, and thus require the policy to compose and transition between the appropriate skills. The goal $\rvg_t = (\tilde{\rvx}^*_t, h_t)$ records the location of the target $\tilde{\rvx}^*_t$ in the character's local coordinate frame, along with an indicator variable $h_t$ that specifies if the target has already been hit.

\begin{figure*}[t!]
	\centering
    \subfigure[Humanoid: Target Location (Locomotion)]{\includegraphics[width=0.498\textwidth]{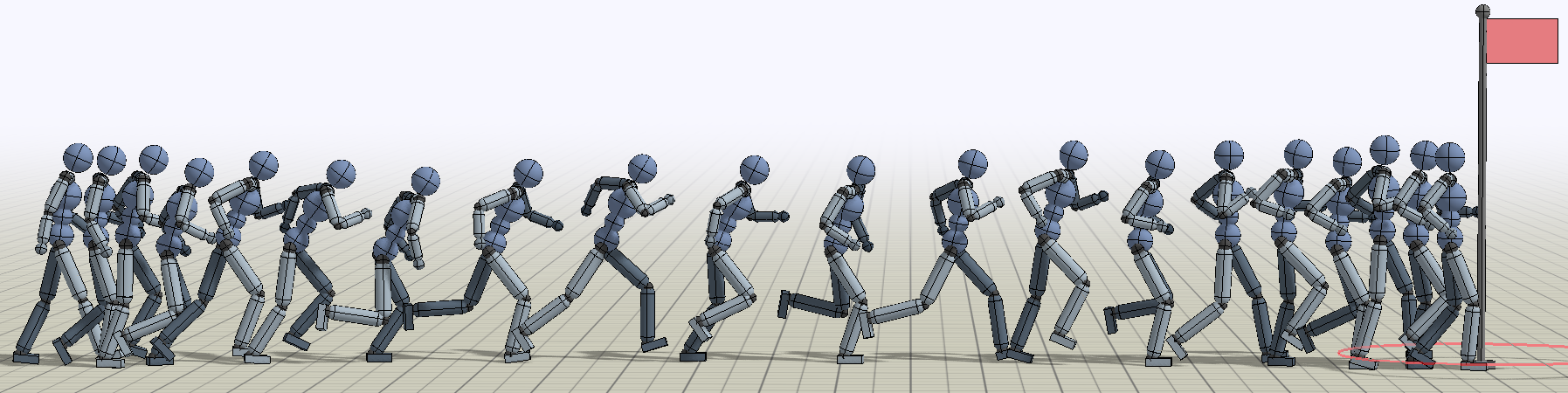}}
    \subfigure[Humanoid: Target Location (Zombie)]{\includegraphics[width=0.498\textwidth]{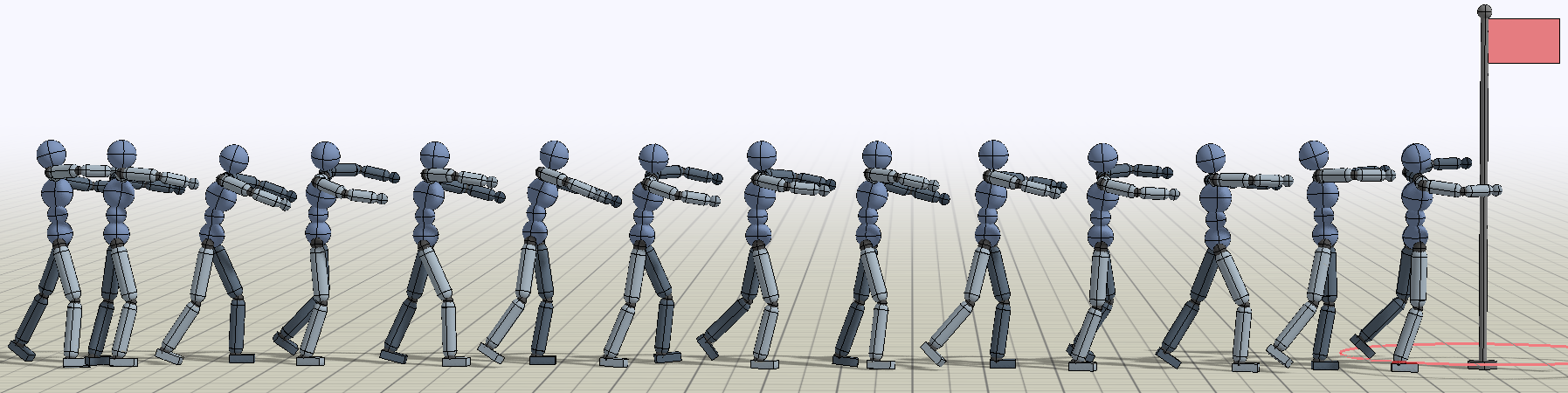}}\\
    \vspace{-0.3cm}
    \subfigure[Humanoid: Target Heading (Locomotion + Getup)]{\includegraphics[width=1\textwidth]{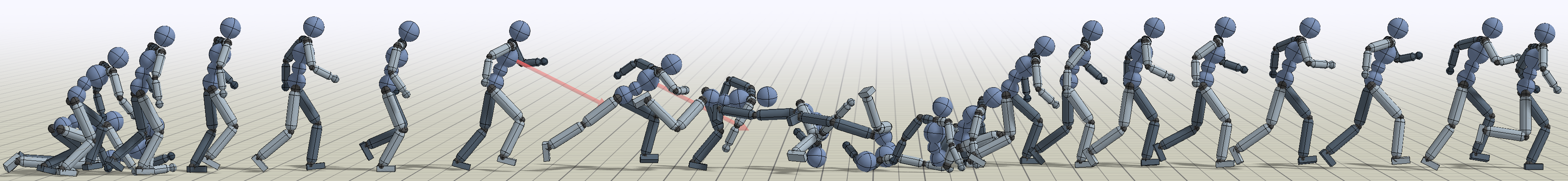} \label{fig:framesTaskGetup}}\\
    \vspace{-0.3cm}
    \subfigure[Humanoid: Dribble (Locomotion)]{\includegraphics[width=0.498\textwidth]{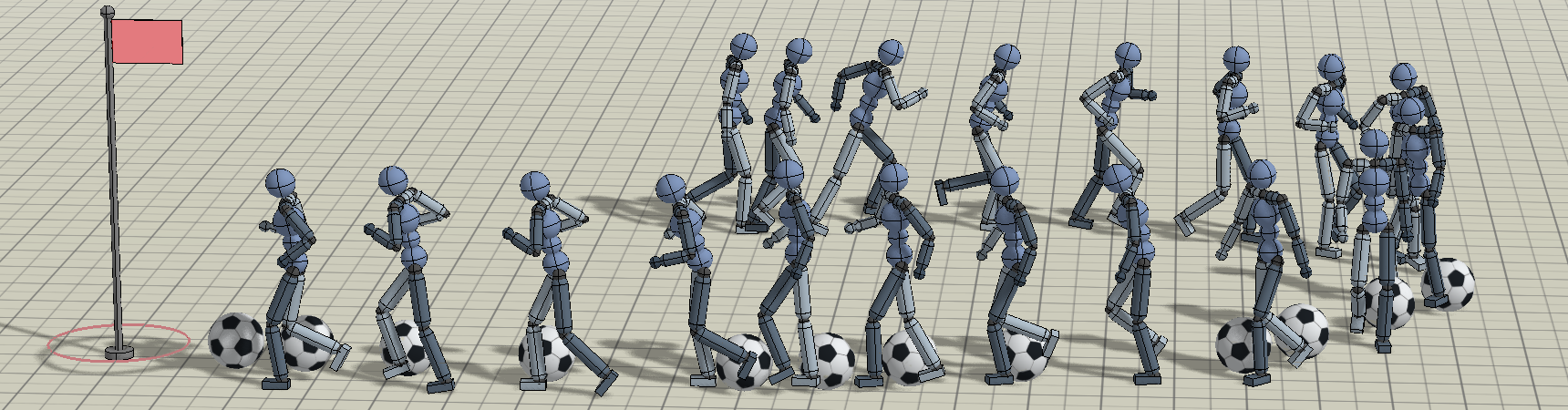}}
    \subfigure[Humanoid: Strike (Walk + Punch)]{\includegraphics[width=0.498\textwidth]{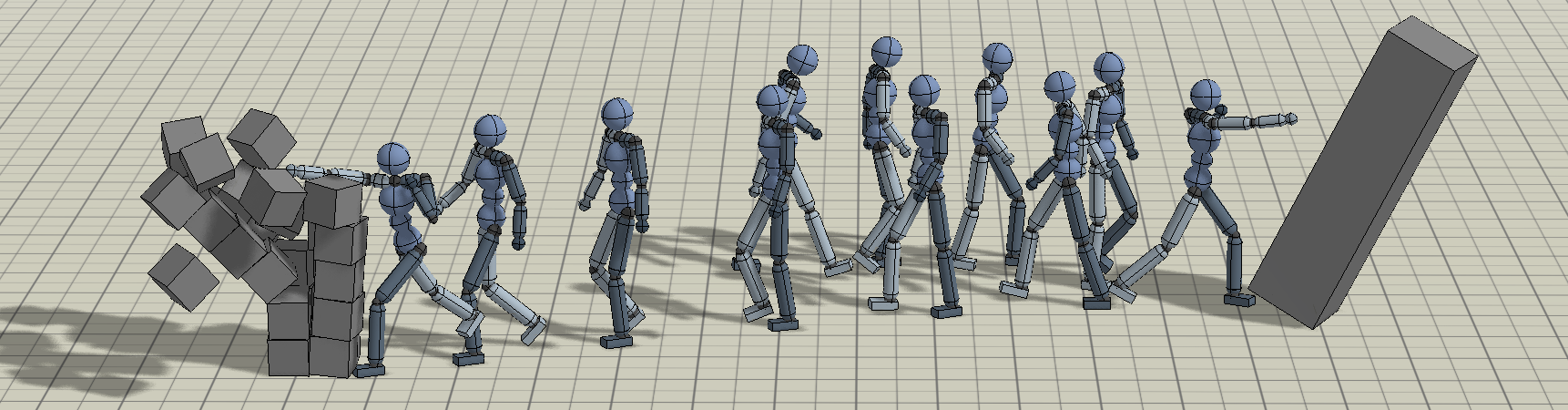}}\\
    \vspace{-0.3cm}
    \subfigure[Humanoid: Obstacles (Run + Leap + Roll)]{\includegraphics[width=1\textwidth]{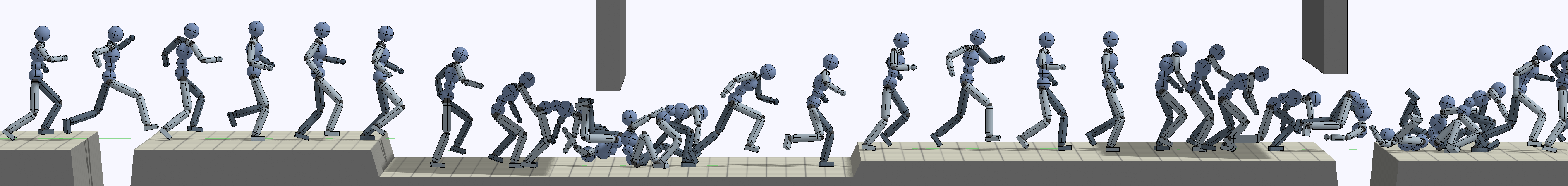} \label{fig:framesObstacles}}\\
    \vspace{-0.3cm}
    \subfigure[Humanoid: Stepping Stones (Cartwheel)]{\includegraphics[width=0.498\textwidth]{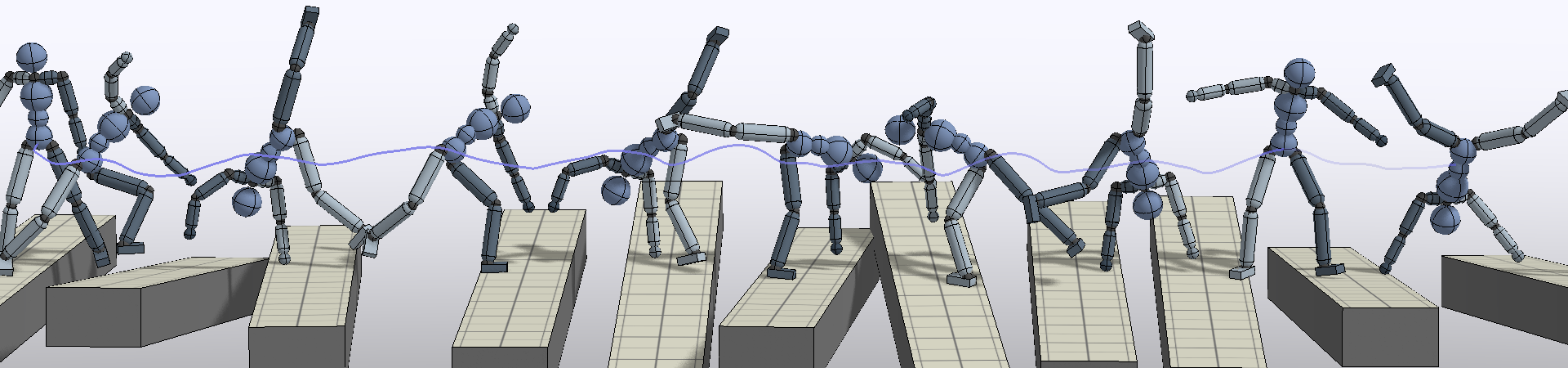}}
    \subfigure[Humanoid: Stepping Stones (Jump)]{\includegraphics[width=0.498\textwidth]{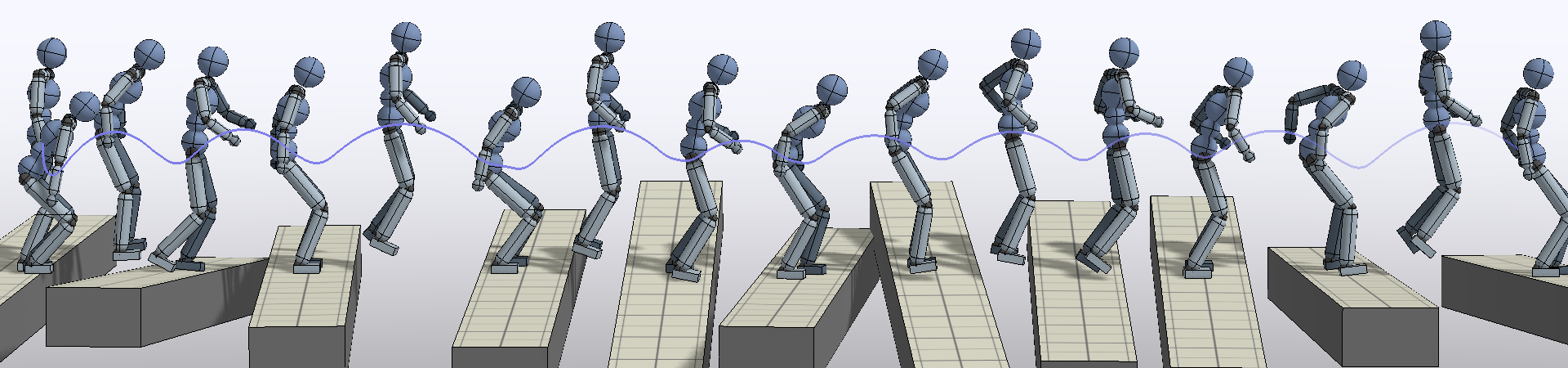}}\\
    \vspace{-0.5cm}
\caption{The motion prior can be trained with large datasets of diverse motions, enabling simulated characters to perform complex tasks by composing a wider range of skills. Each environment is denoted by "Character: Task (Dataset)".}
\label{fig:framesTasks}
\vspace{-0.25cm}
\end{figure*}

\paragraph{Obstacles:}
Finally, we consider tasks that involve visual perception and interaction with more complex environments, where the character's objective is to traverse an obstacle-filled terrain, while maintaining a target speed. Policies are trained for two types of environments: 1) An environment containing a combination of obstacles include gaps, steps, and overhead obstructions that the character must duck under. 2) An environment containing narrow stepping stones that requires more precise contact planning. Examples of the environments are available in Figure~\ref{fig:teaser} and \ref{fig:framesTasks}. In order for the policy to perceive the upcoming obstacles, the state is augmented with a 1D height-field of the upcoming terrain.

\section{Results}
We evaluate our framework's effectiveness on a suite of challenging motion control tasks with complex simulated characters. First, we demonstrate that our approach can readily scale to large unstructured datasets containing diverse motion clips, which then enables our characters to perform challenging tasks in a natural and life-like manner by imitating behaviors from the dataset. The characters automatically learn to compose and generalize different skills from the motion data in order to fulfill high-level task objectives, without requiring mechanisms for explicit motion selection. We then evaluate AMP on a single-clip imitation task, and show that our method is able to closely imitate a diverse corpus of dynamic and acrobatic skills, producing motions that are nearly indistinguishable from reference motions recorded from human actors. Behaviors learned by the characters can be viewed in the supplementary video.

\subsection{Experimental Setup}
All environments are simulated using the Bullet physics engine~\citep{coumans2013bullet}, with a simulation frequency of 1.2kHz. The policy is queried at 30Hz, and each action specifies target positions for PD controllers positioned at the character's joints.
All neural networks are implemented using Tensorflow~\citep{tensorflow2015}. The gradient penalty coefficient is set to $w^\mathrm{gp} = 10$. Detailed hyperparameter settings are available in Appendix~\ref{sec:suppAmpParams}. Reference motion clips are collected from a combination of public mocap libraries, custom recorded mocap clips, and artist-authored keyframe animations \citep{CMUMocap,SFUMocap,MANN2018}. Depending on the task and character, each policy is trained with 100-300 million samples, requiring approximately 30-140 hours on 16 CPU cores. Code for our system will be released upon publication of this paper.

\begin{table}[t]
{ \centering  
\caption{Performance statistics of combining AMP with additional task objectives. Performance is recorded as the average normalized task return, with 0 being the minimum possible return per episode and 1 being the maximum possible return. The return is averaged across 3 models initialized with different random seeds, with 32 episodes recorded per model. The motion prior can be trained with different datasets to produce policies that adopt distinct stylistic behaviors when performing a particular task.}
\vspace{-0.25cm}
\label{tab:taskPerf}
\begin{tabular}{|l|c|c|c|}
\hline
{\bf Character} & {\bf Task} & {\bf Dataset} & {\bf Task Return} \\ \hline
    Humanoid & Target & Locomotion & $0.90 \pm 0.01$ \\ \cline{3-4}
             & Heading & Walk & $0.46 \pm 0.01$   \\ \cline{3-4}
             &                & Run & $0.63 \pm 0.01$  \\ \cline{3-4}
             &                & Stealthy & $0.89 \pm 0.02$  \\ \cline{3-4}
             &                & Zombie & $0.94 \pm 0.00$   \\ \cline{2-4}
             & Target & Locomotion & $0.63 \pm 0.01$   \\ \cline{3-4}
             & Location & Zombie& $0.50 \pm 0.00$   \\ \cline{2-4}
             & Obstacles & Run + Leap + Roll & $0.27 \pm 0.10$ \\ \cline{2-4}
             & Stepping  & Cartwheel & $0.43 \pm 0.03$   \\ \cline{3-4}
             & Stones & Jump& $0.56 \pm 0.12$   \\ \cline{2-4}
             & Dribble & Locomotion & $0.78 \pm 0.05$ \\ \cline{3-4}
             &                           & Zombie & $0.60 \pm 0.04$  \\ \cline{2-4}
             & Strike & Walk + Punch & $0.73 \pm 0.02$  \\ \hline
    \shortstack{T-Rex \\ \ } & \shortstack{Target \\ Location} & \shortstack{Locomotion \\ \ } & \shortstack{$0.36 \pm 0.03$\\ \ }  \\ \hline
\end{tabular} \\
}
\vspace{-0.25cm}
\end{table}

\subsection{Tasks}
In this section, we demonstrate AMP's effectiveness for controlling the style of a character's motions as it performs other high-level tasks. The weights for the task-reward and style-reward are set to $w^G = 0.5$ and $w^S = 0.5$ for all tasks. The character can be trained to perform tasks in a variety of distinct styles by providing the motion prior with different datasets. Figure~\ref{fig:framesTasks} illustrates behaviors learned by the Humanoid on various tasks. Table~\ref{tab:taskPerf} records the performance of the policies with respect to the normalized task return, and summary statistics of the different datasets used to train the motion priors are available in Table~\ref{tab:datasets}. AMP can accommodate large unstructured datasets, with the largest dataset containing 56 clips from 8 different human actors, for a total of 434s of motion data. In the case of the \emph{Target Heading} task, a motion prior trained using a locomotion dataset, containing walking, running, and jogging motions, leads to a policy that executes different locomotion gaits depending on the target speed. Transitions between various gaits emerge automatically through the motion prior, with the character adopting walking gaits at slow speeds ($\sim 1$m/s), switching to jogging gaits at faster speeds ($\sim 2.5$m/s), and breaking into a fast run as the target speed approaches  ($\sim 4.5$m/s). The motion prior also leads to other human-like strategies, such as banking into turns, and slowing down before large changes in direction. The policies develop similar behaviors for the \emph{Target Location} task. When the target is near the character, the policy adopts slower walking gaits. But when the target is further away, the character automatically transitions into a run. 
These intricate behaviors arise naturally from the motion prior, without requiring a motion planner to explicitly select which motion the character should execute in a given scenario, such as those used in prior systems \citep{2017-TOG-deepLoco,DreCon2019,CARLLuo2020}. In addition to standard locomotion gaits, the motion prior can also be trained for more stylistic behaviors, such as walking like a shambling zombie or walking in a stealthy manner. Our framework enables the character to acquire these distinct styles by simply providing the motion prior with different unstructured motion datasets.

\begin{figure}[t]
	\centering
    \subfigure{\includegraphics[height=0.41\columnwidth]{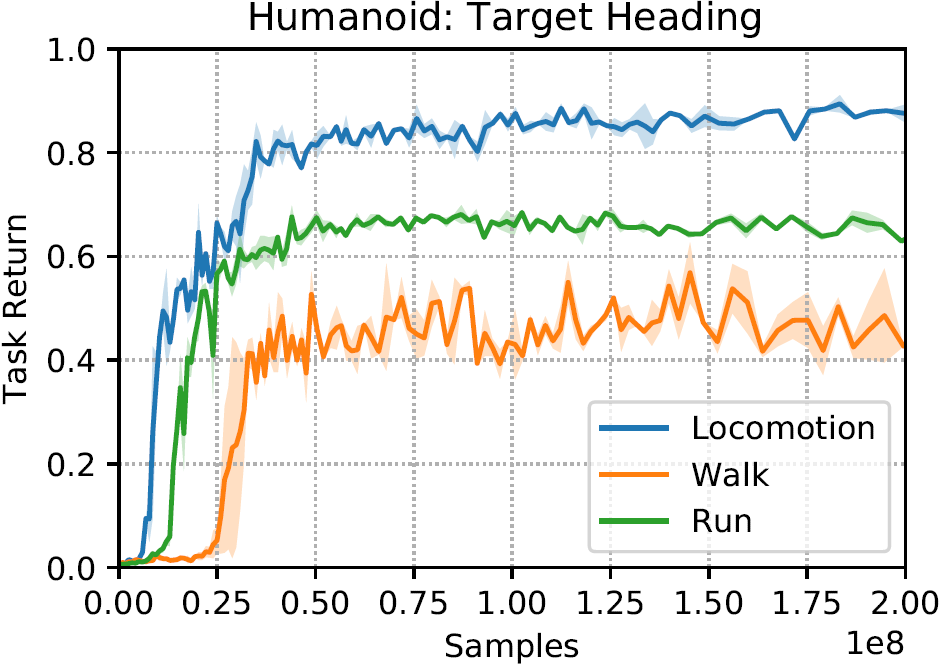}}
    \subfigure{\includegraphics[height=0.41\columnwidth]{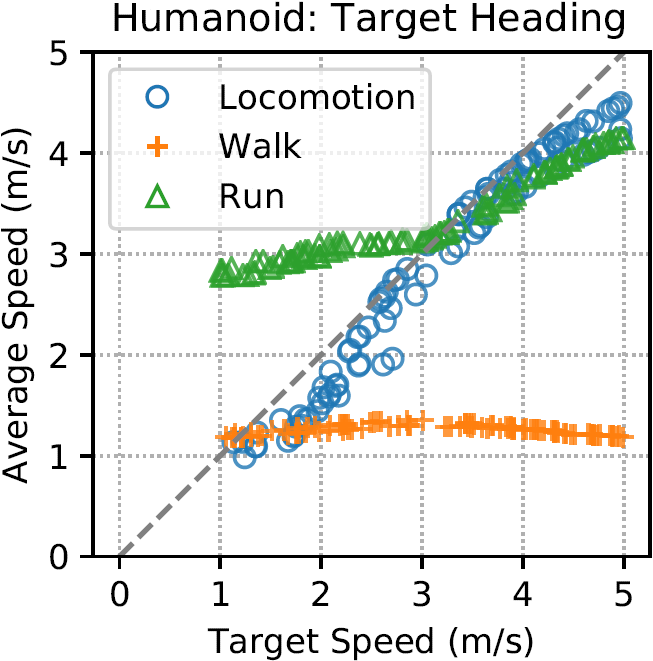}}\\
    \vspace{-0.3cm}
\caption{Performance of Target Heading policies trained with different datasets. \textbf{Left:} Learning curves comparing the normalized task returns of policies trained with a large dataset of diverse locomotion clips to policies trained with only walking or running reference motions. Three models are trained using each dataset. \textbf{Right:} Comparison of the target speed with the average speed achieved by the different policies. Policies trained using the larger Locomotion dataset is able to more closely follow the various target speeds by imitating different gaits.}
\label{fig:targetHeadingAblation}
\end{figure}

To determine whether the transitions between distinct gaits are a product of the motion prior or a result of the task objective, we train policies to perform the \emph{Target Heading} task using limited datasets containing only walking or running data.
Figure~\ref{fig:targetHeadingAblation} compares the performance of policies trained with these different datasets. Policies trained with only walking motions learn to perform only walking gaits, and do not show any transitions to faster running gaits even at faster target speeds. As a result, these policies are not able to achieve the faster target speeds. Similarly, policies trained with only running motions are not able to match slower target speeds. Training the motion prior with a diverse dataset results in more flexible and optimal policies that are able to achieve a wider range of target speeds. This indicates that the diversity of behaviors exhibited by our policies can in large part be attributed to the motion prior, and is not solely a result of the task objective.

\begin{table}[t]
{ \centering  
\caption{Summary statistics of the different datasets used to train the motion priors. We record the total length of motion clips in each dataset, along with the number of clips, and the number of subjects (e.g. human actors) that the clips were recorded from.}
\vspace{-0.25cm}
\label{tab:datasets}
\resizebox{\columnwidth}{!}{
\begin{tabular}{|l|c|c|c|c|}
\hline
{\bf Character} & {\bf Dataset} & {\bf Size (s)} & {\bf Clips} & {\bf Subjects} \\ \hline
    Humanoid & Cartwheel & $13.6$ & $3$  & $1$ \\ \cline{2-5}
             & Jump & $28.6$ & $10$  & $4$ \\ \cline{2-5}
             & Locomotion & $434.1$ & $56$  & $8$ \\ \cline{2-5}
             & Run & $204.4$ & $47$ & $3$ \\ \cline{2-5}
             & Run + Leap + Roll & $22.1$ & $10$  & $7$ \\ \cline{2-5}
             & Stealthy & $136.5$ & $3$  & $1$ \\ \cline{2-5}
             & Walk & $229.6$ & $9$  & $5$ \\ \cline{2-5}
             & Walk + Punch & $247.8$ & $15$  & $9$ \\ \cline{2-5}
             & Zombie & $18.3$ & $1$  & $1$ \\ \hline
    T-Rex & Locomotion & $10.5$ & $5$  & $1$ \\ \hline
\end{tabular}
}
}
\vspace{-0.25cm}
\end{table}

To further illustrate AMP's ability to compose disparate skills, we introduce additional reference motions into the dataset for getting up from the ground in various configurations. These additional motion clips then enable our character to recover from a fall and continue to perform a given task (Figure~\ref{fig:framesTaskGetup}). The policy also discovers novel recovery behaviors that are not present in the dataset. When the character falls forward, it tucks its body into a roll during the fall in order to more quickly transition into a getup behavior.
While this particular behavior is not present in the motion clips, the policy is able to generalize behaviors observed in the dataset to produce novel and naturalistic strategies for new scenarios.

For the \emph{Strike} task (Figure~\ref{fig:teaser}), the motion prior is trained using a collection of walking motion clips and punching motion clips. The resulting policy learns to walk to the target when it is far away, and then transition to a punching motion once it is within range to hit the target. Note that the motion clips in the dataset contain strictly walking-only motions or punching-only motion, and none of the clips show an actor walking to and punching a target. Instead, the policy learns to temporally sequence these different behaviors in order to fulfill the high-level task objectives. Again, this composition of different skills emerges automatically from the motion prior, without requiring a motion planner or other mechanisms for motion selection.

\begin{figure}[t]
	\centering
    \subfigure{\includegraphics[height=0.365\columnwidth]{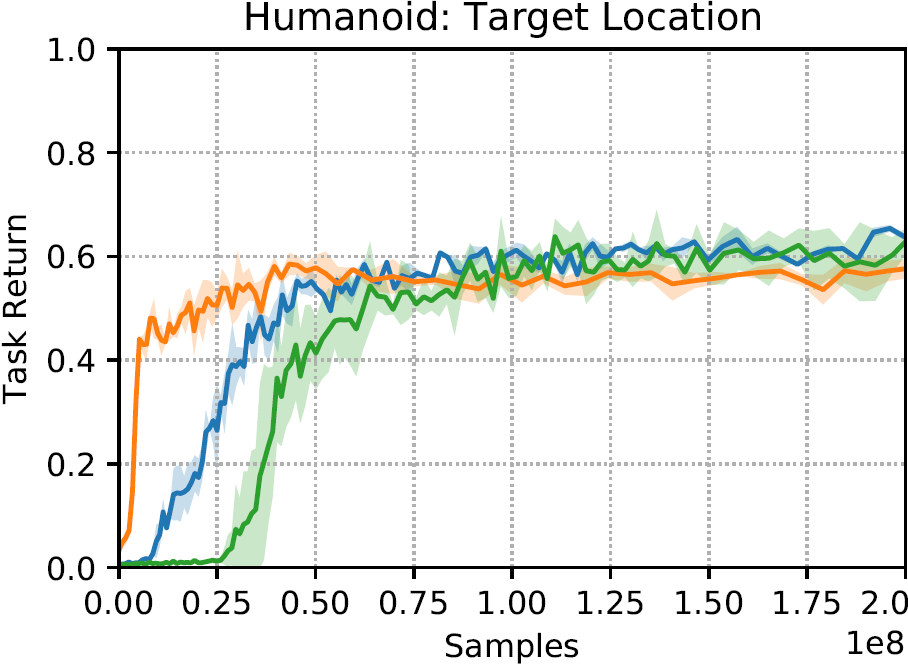}}
    \subfigure{\includegraphics[height=0.365\columnwidth]{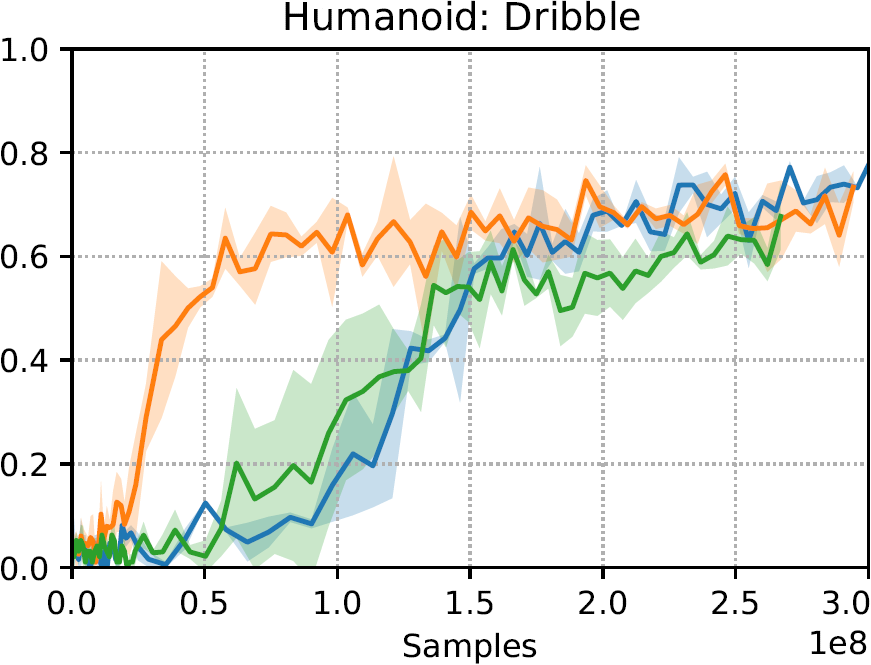}}\\
    \vspace{-0.1cm}
    \subfigure{\includegraphics[height=0.365\columnwidth]{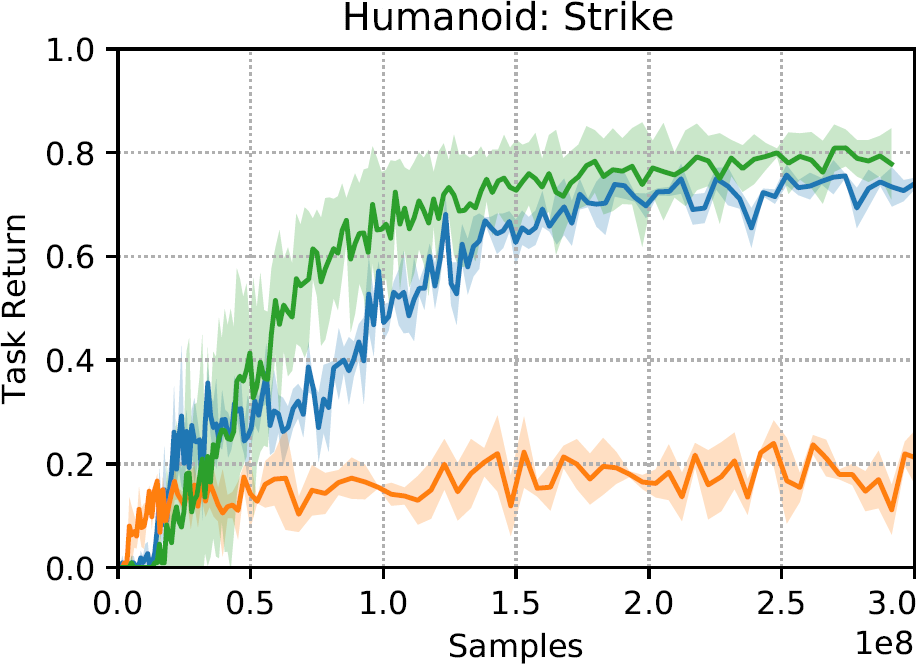}} 
    \subfigure{\includegraphics[height=0.365\columnwidth]{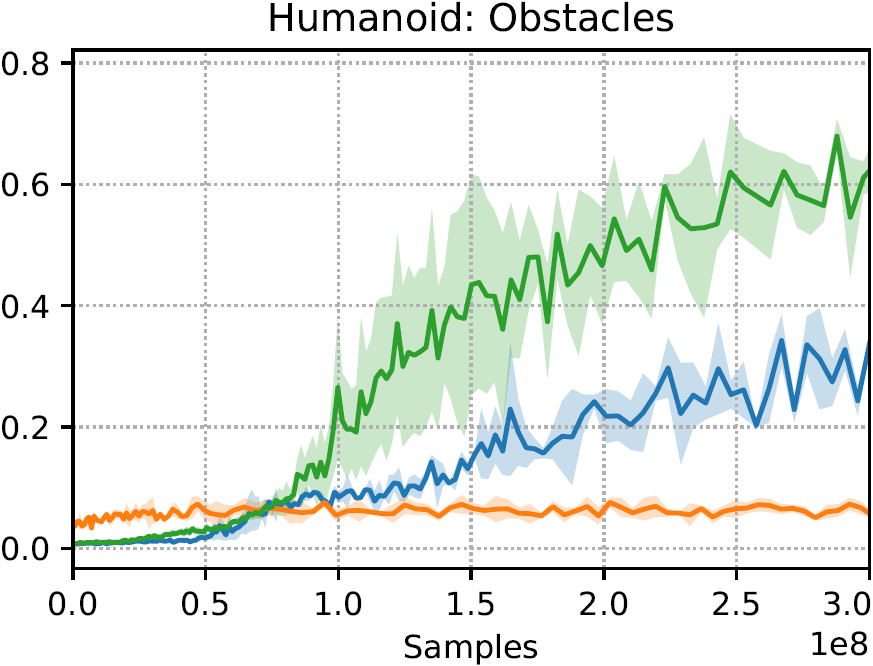}}\\
    \vspace{-0.1cm}
    \subfigure{\includegraphics[height=0.027\columnwidth]{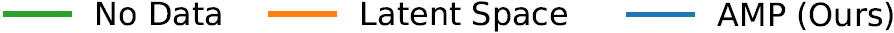}}\\
    \vspace{-0.3cm}
\caption{Learning curves comparing the task performance of AMP to latent space models (Latent Space) and policies trained from scratch without motion data (No Data). Our method achieves comparable performance across the various tasks, while also producing higher fidelity motions.} 
\label{fig:curvesComp}
\vspace{-0.25cm}
\end{figure}

Finally, our system can also train visuomotor policies for traversing obstacle-filled environments. By providing the motion prior with a collection of locomotion clips and rolling clips, the character learns to utilize these diverse behaviors to traverse the different obstacles. The character learns to leap over obstacles such as gaps. But as it approaches the overhead obstructions, the character transitions into a rolling behavior in order to pass underneath the obstacles. Previous systems that have demonstrated similar composition of diverse maneuvers for clearing obstacle have typically required a separate motion planner or manual annotations \citep{2012-TOG-TerrainRunner,PredictSimPark2019}. Our approach provides a unified framework where the same underlying algorithm is able to learn how to perform the various skills and which skill to execute in a given scenario. Furthermore, the character can also be trained to traverse obstacles in distinct styles by providing the motion prior with different motion clips, such as jumping or cartwheeling across stepping stones (Figure~\ref{fig:framesTasks}).

\subsection{Comparisons}
An alternative approach for learning a motion prior from unstructured motion data is to build a latent space model \citep{HeessWTLRS16,lynch20a,MCPPeng19,CatchCarryMerel2020}.
Unlike AMP, which encourages a character to adopt a desired motion style directly through an optimization objective, a latent space model enforces a particular motion style indirectly, by using a latent representation to constrain the policy's actions to those that produce motions of the desired style. To compare AMP to these latent space models, we first pre-train a low-level controller using a motion tracking objective to imitate the same set of reference motions that are used to train the motion prior. The learned low-level controller is then used to train separate high-level controllers for each downstream task. Note that reference motions are used only during pre-training, and the high-level controllers are trained to optimize only the task objectives. A more in-depth description of the experimental setup for the latent space model is available in Appendix~\ref{sec:suppLatentSpace}.

A qualitative comparison of the behaviors learned using AMP and the latent space model is available in the supplementary video. Figure~\ref{fig:curvesComp}
compares the task performance of the different models, along with a baseline model trained from scratch for each task without leveraging any motion data. Both AMP and the latent space models are able to produce substantially more life-like behaviors than the baseline models. For the latent space models, since the low-level and high-level controllers are trained separately, it is possible for the distribution of encodings specified by the high-level controller to be different than the distribution of encodings observed by the low-level controller during pre-training \citep{CARLLuo2020}. This in turn can result in unnatural motions that deviate from the behaviors depicted in the original dataset. AMP enforces a motion style directly through the reward function, and is therefore able to better mitigate some of these artifacts. The more structured exploration behaviors from the latent space model enable the policies to solve downstream tasks more quickly. However, the pre-training stage used to construct the low-level controller can itself be sample intensive. In our experiments, the low-level controllers are trained using 300 million samples before being transferred to downstream tasks. With AMP, no such pre-training is necessary, and the motion prior can be trained jointly with the policy.

\begin{figure}[t]
	\centering
    \subfigure{\includegraphics[width=1\columnwidth]{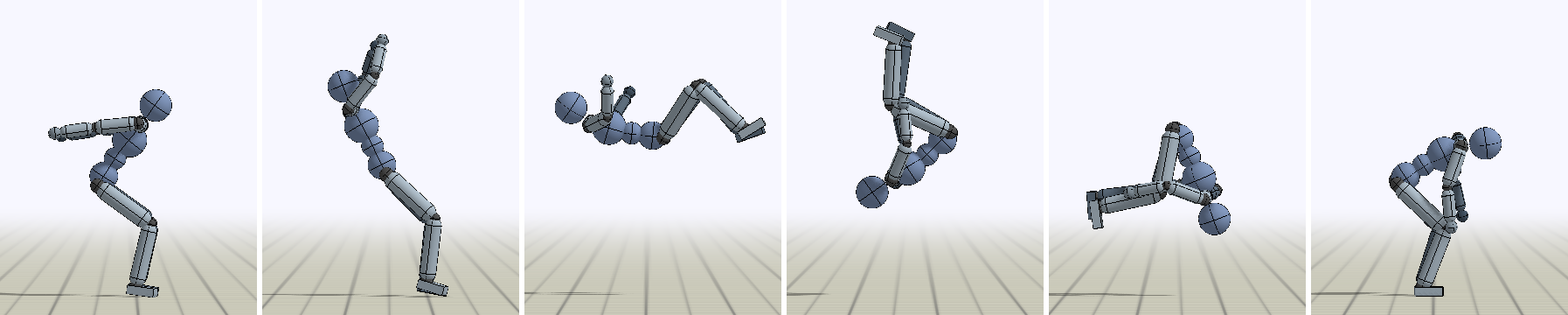}}\\
    \vspace{-0.3cm}
    \subfigure{\includegraphics[width=1\columnwidth]{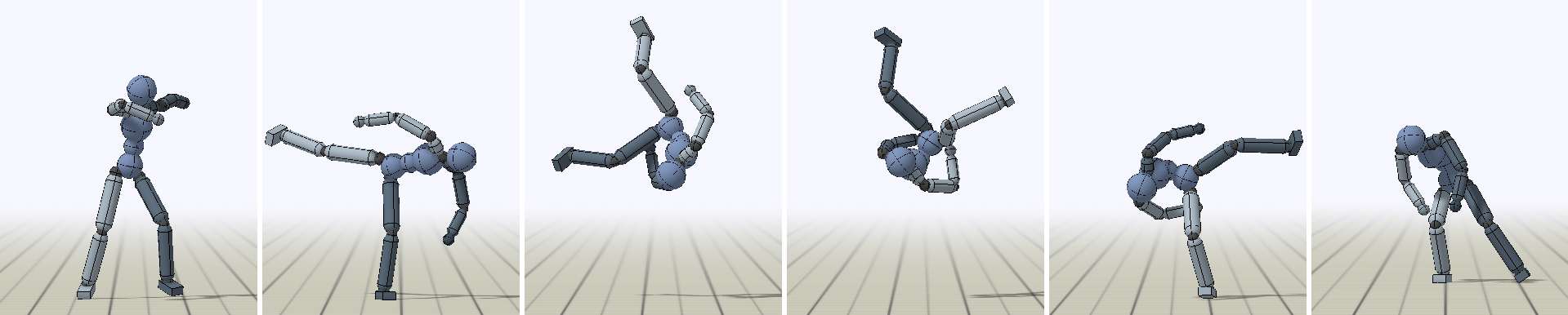}}\\
    \vspace{-0.3cm}
    \subfigure{\includegraphics[width=1\columnwidth]{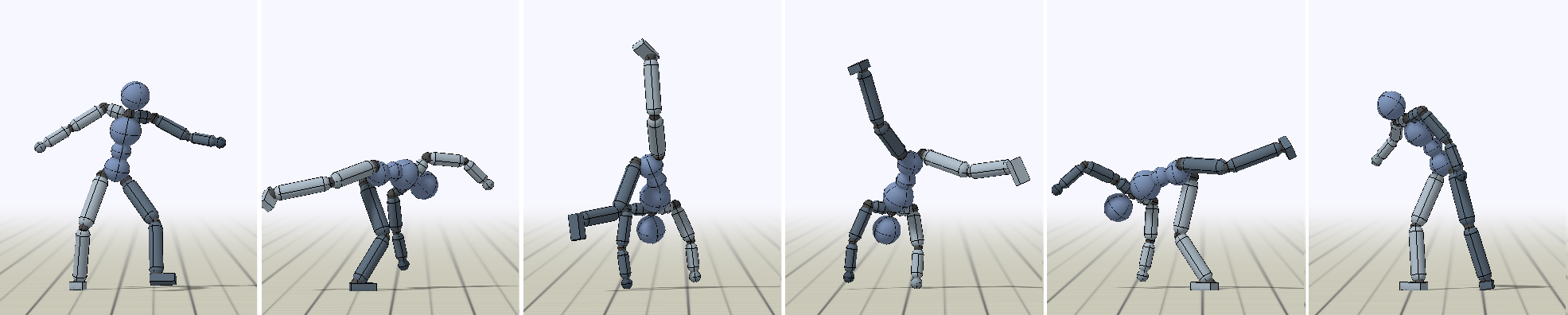}}\\
    \vspace{-0.3cm}
    \subfigure{\includegraphics[width=1\columnwidth]{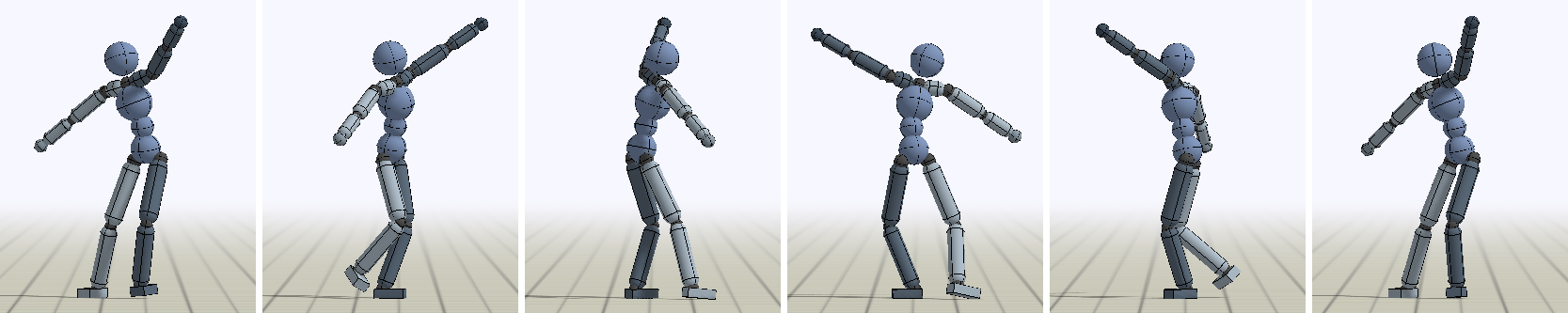}}\\
    \subfigure{\includegraphics[width=1\columnwidth]{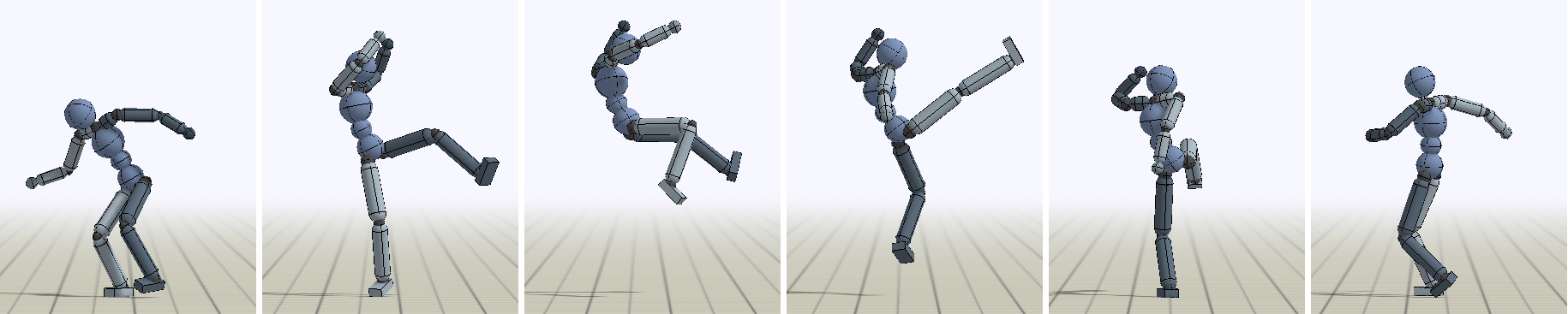}}\\
    \vspace{-0.3cm}
    \subfigure{\includegraphics[width=1\columnwidth]{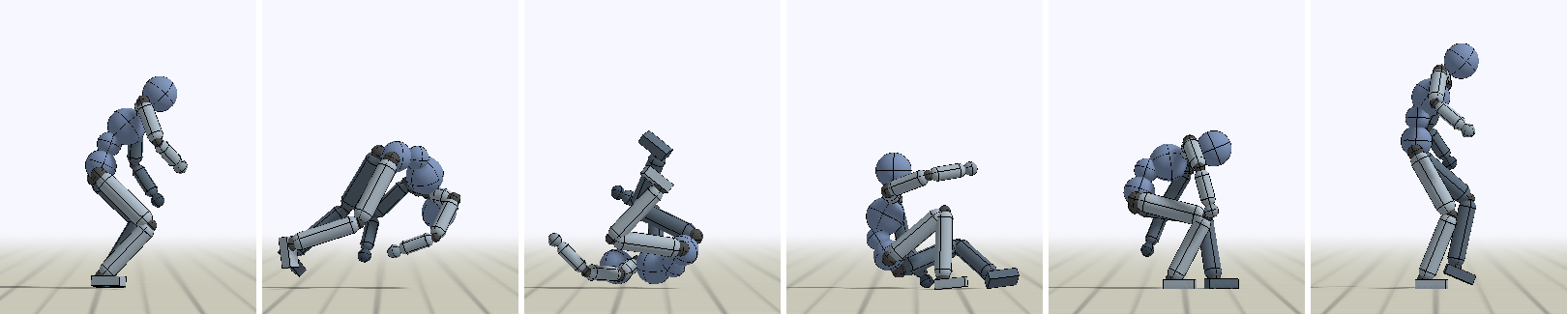}}\\
    \vspace{-0.4cm}
\caption{Snapshots of behaviors learned by the Humanoid on the single-clip imitation tasks. \textbf{Top-to-bottom: } back-flip, side-flip, cartwheel, spin, spin-kick, roll. AMP enables the character to closely imitate a diverse corpus of highly dynamic and acrobatic skills.}
\label{fig:framesHumanoid}
\vspace{-0.4cm}
\end{figure}

\begin{figure*}[t!]
	\centering
    \subfigure[T-Rex (Walk)]{\includegraphics[width=1\textwidth]{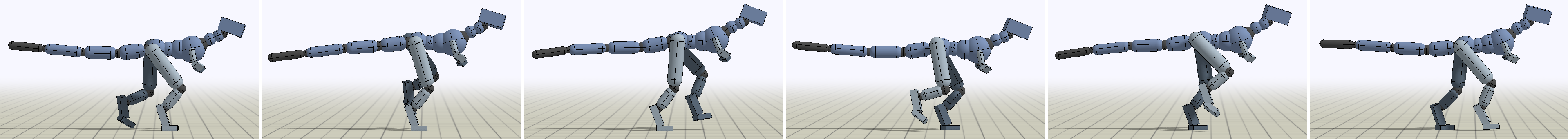}}\\
    \vspace{-0.3cm}
    \subfigure[Dog (Trot)]{\includegraphics[width=1\textwidth]{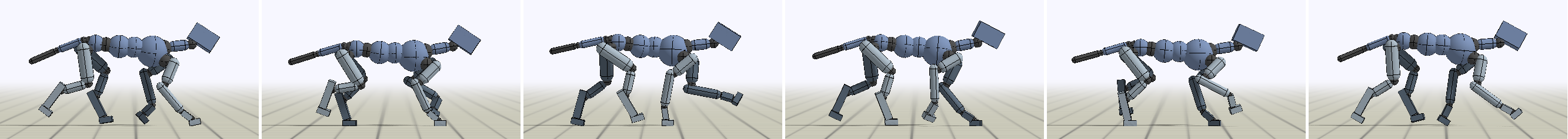}}\\
    \vspace{-0.3cm}
    \subfigure[Dog (Canter)]{\includegraphics[width=1\textwidth]{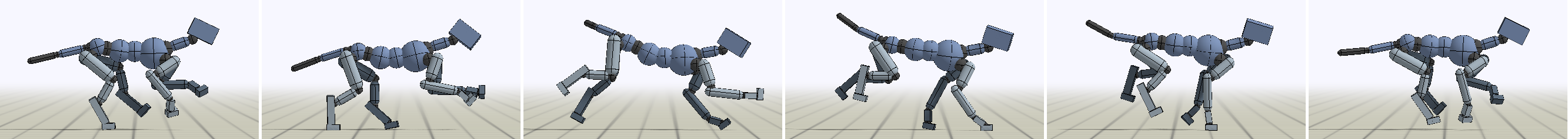}}\\
    \vspace{-0.25cm}
\caption{AMP can be used to train complex non-humanoid characters, such as a 59 DoF T-Rex and a 64 DoF dog. By providing the motion prior with different reference motion clips, the characters can be trained to perform various locomotion gaits, such as trotting and cantering.}
\label{fig:framesTrexDog}
\vspace{-0.1cm}
\end{figure*}

\subsection{Single-Clip Imitation}
Although our goal is to train characters with large motion datasets, to evaluate the effectiveness of our framework for imitating behaviors from motion clips, we consider a single-clip imitation task. In this setting, the character's objective is to imitate a single motion clip at a time, without additional task objectives. Therefore, the policy is trained solely to maximize the style-reward $r_t^S$ from the motion prior. Unlike previous motion tracking methods, our approach does not require a manually designed tracking objective or a phase-based synchronization of the reference motion and the policy \citet{2018-TOG-deepMimic}.
Table~\ref{tab:imitationPerf} summarizes the performance of policies trained using AMP to imitate a diverse corpus of motions. Figure~\ref{fig:framesHumanoid} and \ref{fig:framesTrexDog} illustrate examples of motions learned by the characters. Performance is evaluated using the average pose error, where the pose error $e^\mathrm{pose}_t$ at each time step $t$ is computed between the pose of the simulated character and the reference motion using the relative positions of each joint with respect to the root (in units of meters),
\begin{equation}
    e^\mathrm{pose}_t = \frac{1}{N^\mathrm{joint}} \sum_{j \in \mathrm{joints}} \left|\left|(\rvx^j_t - \rvx^\mathrm{root}_t) - (\hat{\rvx}^j_t - \hat{\rvx}^\mathrm{root}_t)\right|\right|_2 .
\label{eqn:poseErr}
\end{equation}
$\rvx^j_t$ and $\hat{\rvx}^j_t$ denote the 3D Cartesian position of joint $j$ from the simulated character and the reference motion, and $N^\mathrm{joint}$ is the total number of joints in the character's body. This method of evaluating motion similarity
has previously been reported to better conform to human perception of motion similarity \citep{Harada2004,MotionSimTang2008}. 
Since AMP does not use a phase variable to synchronize the policy with the reference motion,
the motions may progress at different rates, resulting in de-synchronization that can lead to large pose errors even when the overall motions are similar. To better evaluate the similarity of the motions, we first apply dynamic time warping (DTW) to align the reference motion with the motion of the simulated character \citep{DTWSakoe1978}, before computing the pose error between the two aligned motions. DTW is applied using Equation~\ref{eqn:poseErr} as the cost function.

\begin{figure}[t]
	\centering
    \subfigure{\includegraphics[height=0.365\columnwidth]{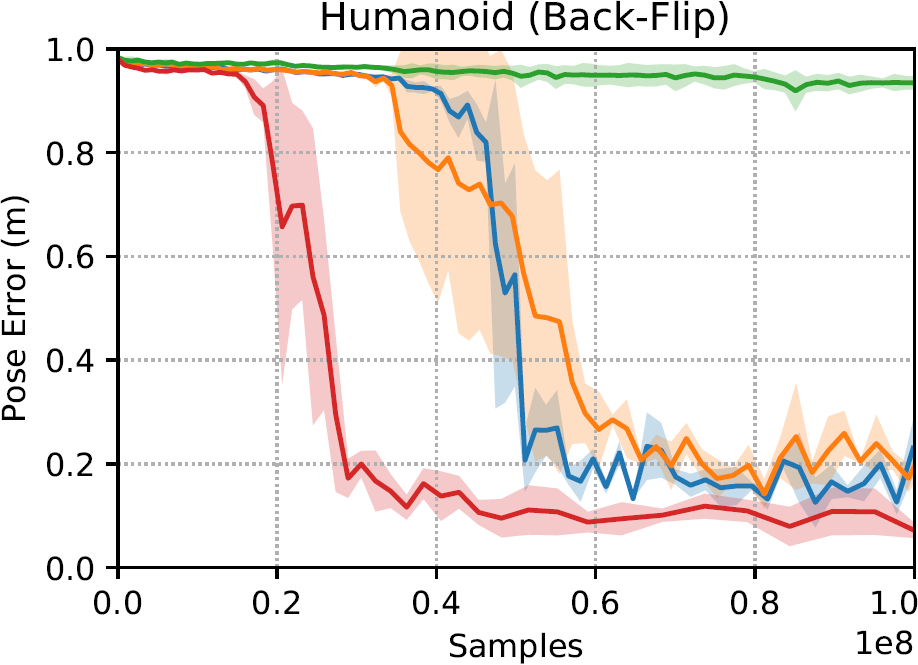}}
    \subfigure{\includegraphics[height=0.365\columnwidth]{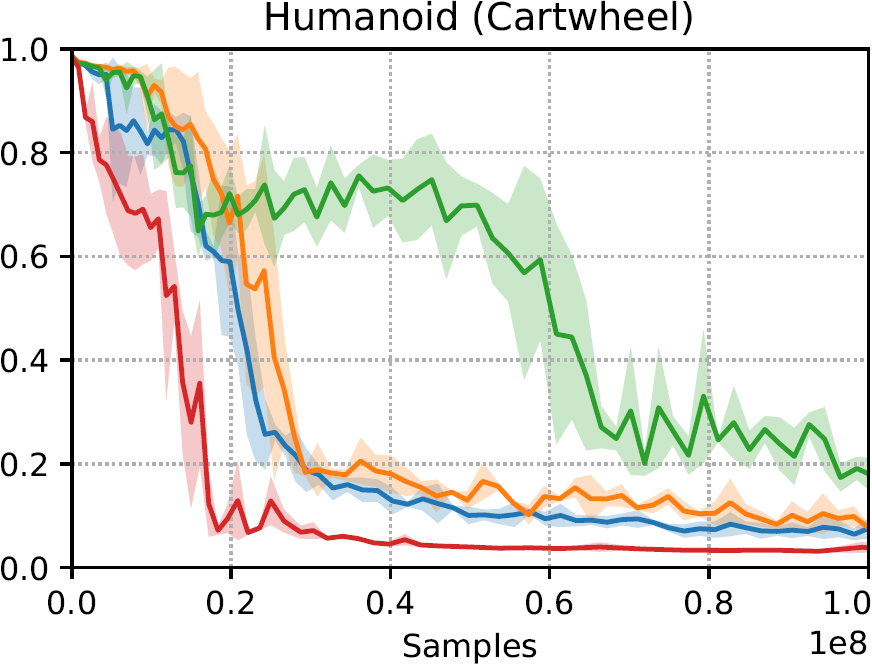}}\\
    \vspace{-0.1cm}
    \subfigure{\includegraphics[height=0.365\columnwidth]{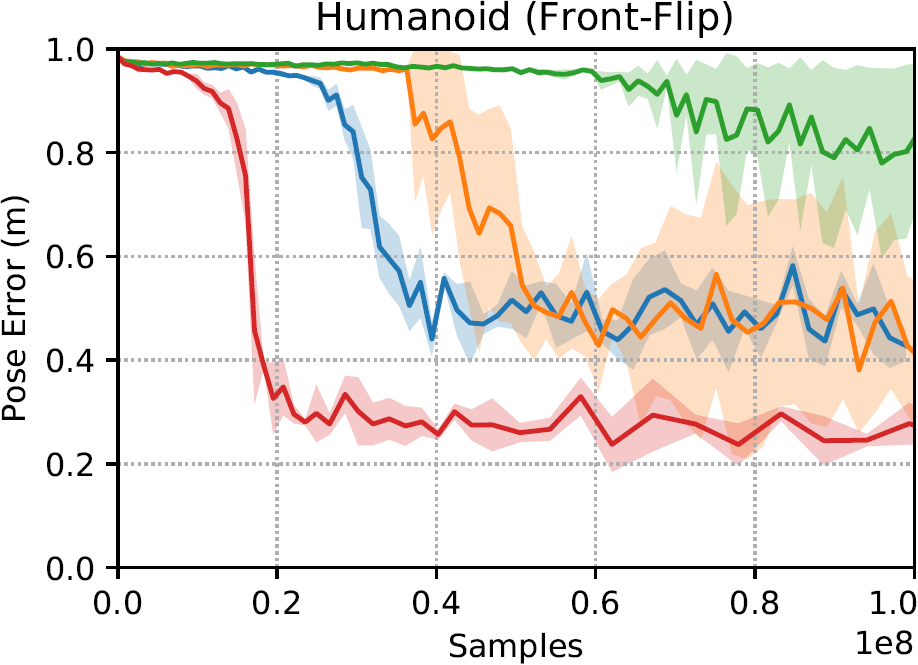}}
    \subfigure{\includegraphics[height=0.365\columnwidth]{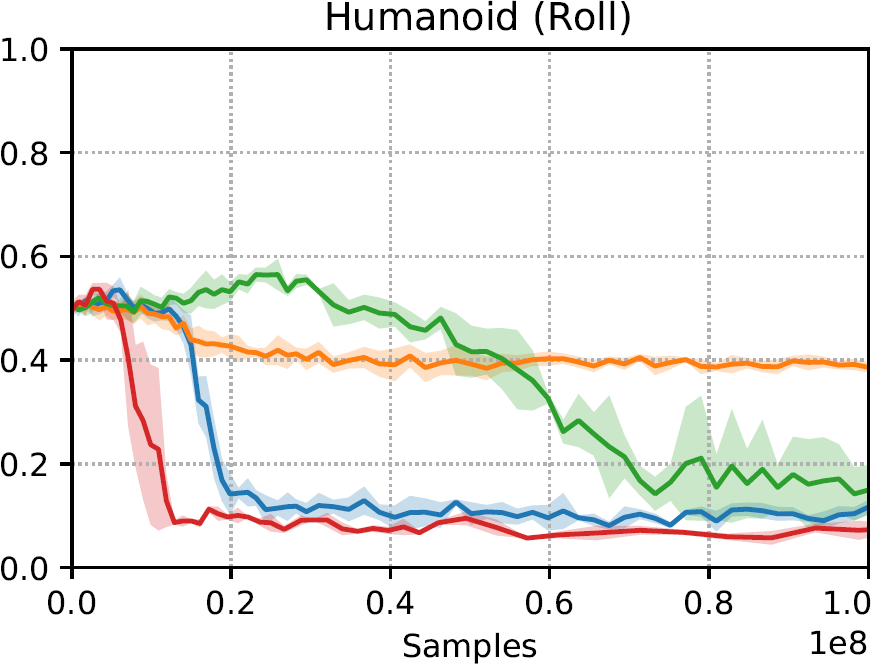}}\\
    \subfigure{\includegraphics[height=0.027\columnwidth]{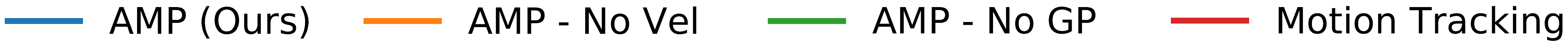}}\\
    \vspace{-0.25cm}
\caption{Learning curves of various methods on the single-clip imitation tasks. We compare AMP to the motion tracking approach proposed by \citet{2018-TOG-deepMimic} (Motion Tracking), as well a version of AMP without velocity features for the discriminator (AMP - No Vel), and AMP without the gradient penalty regularizer (AMP - No GP). A comprehensive collection of learning curves for all skills are available in the Appendix. AMP produces results of comparable quality when compared to prior tracking-based methods, without requiring a manually designed reward function or synchronization between the policy and reference motion.
Velocity features and gradient penalty are vital for effective and consistent results on challenging skills.}
\label{fig:curvesImitation}
\vspace{-0.25cm}
\end{figure}

AMP is able to closely imitate a large variety of highly dynamic skills, while also avoiding many of the visual artifacts exhibited by prior adversarial motion imitation systems \citep{Merel2017,DiverseImitationWang2017}. We compare the performance of our system to results produced by the motion tracking approach from \citet{2018-TOG-deepMimic}, which uses a manually designed reward function and requires synchronization of the policy with a reference motion via a phase variable. Figure~\ref{fig:curvesImitation} compares the learning curves of the different methods. Since the tracking-based policies are synchronized with their respective reference motions, they are generally able to learn faster and achieve lower errors than policies trained with AMP. Nonetheless, our method is able to produce results of comparable quality without the need to manually design or tune reward functions for different motions. However, for some motions, such as the Front-Flip, AMP is prone to converging to locally optimal behaviors, where instead of performing a flip, the character learns to simply shuffle forwards in order to avoid falling. Tracking-based methods can mitigate these local optima by terminating an episode early if the character's pose deviates too far from the reference motion \citep{2018-TOG-deepMimic,Won2020}. However, this strategy is not directly applicable to AMP, since the policy is not synchronized with the reference motion. But as shown in the previous sections, this lack of synchronization is precisely what allows AMP to easily leverage large datasets of diverse motion clips to solve more complex tasks.

\begin{table}[t]
{ \centering  
\caption{Performance statistics of imitating individual motion clips without task objectives. "Dataset Size" records the total length of motion data used for each skill. Performance is recorded as the average pose error (in units of meters) between the time-warped trajectories from the reference motion and simulated character. The pose error is averaged across 3 models initialized with different random seeds, with 32 episodes recorded per model. Each episode has a maximum length of 20s. We compare our method (AMP) with the motion tracking approach proposed by \citet{2018-TOG-deepMimic}. AMP is able to closely imitate a diverse repertoire of complex motions, without manual reward engineering.}
\vspace{-0.25cm}
\label{tab:imitationPerf}
\resizebox{\columnwidth}{!}{
\begin{tabular}{|l|c|c|c|c|}
\hline
{\bf \shortstack{Character \\ \ }} & {\bf \shortstack{Motion \\ \ }} & {\bf \shortstack{Dataset \\ Size}} & {\bf \shortstack{Motion \\ Tracking}} & {\bf \shortstack{AMP \\ (Ours)}} \\ \hline
    Humanoid & Back-Flip & $1.75$s & $ 0.076 \pm 0.021 $  & $0.150 \pm 0.028$ \\ \cline{2-5}
             & Cartwheel & $2.72$s & $ 0.039 \pm 0.011 $  & $0.067 \pm 0.014$ \\ \cline{2-5}
             & Crawl & $2.93$s & $ 0.044 \pm 0.001 $  & $0.049 \pm 0.007$ \\ \cline{2-5}
             & Dance & $1.62$s & $ 0.038 \pm 0.001 $  & $0.055 \pm 0.015$ \\ \cline{2-5}
             & Front-Flip & $1.65$s & $ 0.278 \pm 0.040 $  & $0.425 \pm 0.010$ \\ \cline{2-5}
             & Jog & $0.83$s & $ 0.029 \pm 0.001$  & $0.056 \pm 0.001$ \\ \cline{2-5}
             & Jump & $1.77$s & $ 0.033 \pm 0.001 $  & $0.083 \pm 0.022$ \\ \cline{2-5}
             & Roll & $2.02$s & $ 0.072 \pm 0.018 $  & $0.088 \pm 0.008$ \\ \cline{2-5}
             & Run & $0.80$s & $ 0.028 \pm 0.002 $  & $	0.075 \pm 0.015$ \\ \cline{2-5}
             & Spin & $1.00$s & $  0.063 \pm 0.022 $  & $0.047 \pm 0.002$ \\ \cline{2-5}
             & Side-Flip & $2.44$s & $  0.191 \pm 0.043$  & $0.124 \pm 0.012$ \\ \cline{2-5}
             & Spin-Kick & $1.28$s & $ 0.042 \pm 0.001 $  & $0.058 \pm 0.012$ \\ \cline{2-5}
             & Walk & $1.30$s & $ 0.018 \pm 0.005$  & $0.030 \pm 0.001$ \\ \cline{2-5}
             & Zombie & $1.68$s & $ 0.049 \pm 0.013 $  & $0.058 \pm 0.014$ \\ \hline
    T-Rex & Turn & $2.13$s & $0.098 \pm 0.011$  & $0.284 \pm 0.023$ \\ \cline{2-5}
          & Walk & $2.00$s & $0.069 \pm 0.005$  & $0.096 \pm 0.027$ \\ \hline
    Dog & Canter & $0.45$s & $0.026 \pm 0.002$  & $0.034 \pm 0.002$ \\ \cline{2-5}
        & Pace & $0.63$s & $0.020 \pm 0.001$  & $0.024 \pm 0.003$ \\ \cline{2-5}
        & Spin & $0.73$s & $0.026 \pm 0.002$  & $0.086 \pm 0.008$ \\ \cline{2-5}
        & Trot & $0.52$s & $0.019 \pm 0.001$  & $0.026 \pm 0.001$ \\ \hline
\end{tabular}
}
}
\vspace{-0.25cm}
\end{table}

\subsection{Ablations}

Our system is able to produce substantially higher fidelity motions
than prior adversarial learning frameworks for physics-based character control \citep{Merel2017,DiverseImitationWang2017}. In this section, we identify critical design decisions that lead to more stable training and higher quality results. Figure~\ref{fig:curvesImitation} compares learning curves of policies trained on the single-clip imitation tasks with different components of the system disabled. Gradient penalty proves to be the most vital component. Models trained without this regularization tend to exhibit large performance fluctuations over the course of the training, and lead to noticeable visual artifacts in the final policies, as shown in the supplementary video. The addition of the gradient penalty not only improves stability during training, but also leads to substantially faster learning across a large set of skills. The inclusion of velocity features in the discriminator's observations is also an important component for imitating some motions. In principle, including consecutive poses as input to the discriminator should provide some information that can be used to infer the joint velocities. But we found that this was insufficient for some motions, such as rolling. As shown in the supplementary video, in the absence of velocity features, the character is prone to converging to a strategy of holding a fixed pose on the ground, instead of performing a roll. The additional velocity features are able to mitigate these undesirable behaviors.

\section{Discussion and Limitations}
In this work, we presented an adversarial learning system for physics-based character animation that enables characters to imitate diverse behaviors from large unstructured datasets, without the need for motion planners or other mechanisms for clip selection. Our system allows users to specify high-level task objectives for controlling a character's behaviors, while the more granular low-level style of a character's motions can be controlled using a learned motion prior. Composition of disparate skills in furtherance of a task objective emerges automatically from the motion prior.
The motion prior also enables our characters to closely imitate a rich repertoire of highly dynamic skills, and produces results that are on par with tracking-based techniques, without requiring manual reward engineering or synchronization between the controller and the reference motions.

Our system demonstrates that adversarial imitation learning techniques can indeed produce high fidelity motions for complex skills. However, like many other GAN-based techniques, AMP is susceptible to mode collapse. When provided with a large dataset of diverse motion clips, the policy is prone to imitating only a small subset of the example behaviors, ignoring other behaviors that may ultimately be more optimal for a given task.
The motion priors in our experiments are also trained from scratch for each policy. But since the motion prior is largely task-agnostic, it should in principle be possible to transfer and reuse motion priors for different policies and tasks. Exploring techniques for developing general and transferable motion priors may lead to modular objective functions that can be conveniently incorporated into downstream tasks, without requiring retraining for each new task. While the motion prior does not require direct access to task-specific information, the data used to train the motion prior is generated by policies trained to perform a particular task. This may introduce some task dependencies into the motion prior, which can hinder its ability to be transferred to other tasks. Training motion priors using data generated from larger and more diverse repertoires of tasks may help to facilitate transferring the learned motion priors to new tasks. Our experiments also focus primarily on tasks that involve temporal composition of different skills, which require the character to perform different behaviors at different points in time. However, spatial composition might also be vital for some tasks that require a character to perform multiple skills simultaneously. Developing motion priors that are more amenable to spatial composition of disparate skills may lead to more flexible and sophisticated behaviors. Despite these limitations, we hope this work provides a useful tool that enables physically simulated characters to take advantage of the large motion datasets that have been so effective for kinematic animation techniques, and open exciting directions for future exploration in data-driven physics-based character animation.

\section*{Acknowledgements}
We thank Sony Interactive Entertainment for providing reference motion data for this project, Bonny Ho for narrating the video, the anonymous reviewers for their helpful feedback, and AWS for providing computational resources. This research was funded by an NSERC Postgraduate Scholarship, and a Berkeley Fellowship for Graduate Study.

\bibliographystyle{ACM-Reference-Format}
\bibliography{motion_prior}

\clearpage

\section*{Appendix}

\appendix

\section{Tasks}
\label{supp:tasks}

In this section, we provide a detailed description of each task, and the task reward functions used during training.

\paragraph{Target Heading:}
In this task, the objective for the character is to move along a target heading direction $\rvd^*$ at a target speed $v^*$. The goal input for the policy is specified as $\rvg_t = (\tilde{\rvd}^*_t, v^*)$, with $\tilde{\rvd}^*_t$ being the target direction in the character's local coordinate frame. The task-reward is calculated according to:
\begin{equation}
    r^G_t = \mathrm{exp}\left(-0.25 \left(v^* - \rvd^* \cdot \dot{\rvx}_t^\mathrm{com} \right)^2 \right),
\label{eqn:rewardHeading}
\end{equation}
where $\dot{x}_t^\mathrm{com}$ is the center-of-mass velocity of the character at time step $t$, and the target speed is selected randomly between $v^* \in [1, 5]$m/s. For slower moving styles, such as Zombie and Stealthy, the target speed is fixed at $1$m/s.

\paragraph{Target Location:} In this task, the character's objective is to move to a target location $\rvx^*$. The goal $\rvg_t = \tilde{\rvx}^*_t$ records the target location in the character's local coordinate frame. The task-reward is given by:
\begin{align}
    r^G_t & = 0.7 \ \mathrm{exp}\left(-0.5 ||\rvx^* - \rvx^\mathrm{root}_t||^2 \right) \nonumber \\ 
    & + 0.3 \ \mathrm{exp}\left(- \left(\mathrm{max}\left(0, \ v^* - \rvd^*_t \cdot \dot{\rvx}_t^\mathrm{com} \right)\right)^2 \right).
\label{eqn:rewardLocation}
\end{align}
Here, $v^* = 1m/s$ specifies a minimum target speed at which the character should move towards the target, and the character will not be penalized for moving faster than this threshold. $\rvd^*_t$ is a unit vector on the horizontal plane that points from the character's root to the target.

\paragraph{Dribbling:}
To evaluate our system on more complex object manipulation tasks, we train policies for a dribbling task, where the objective is for the character to dribble a soccer ball to a target location. The reward function is given by:
\begin{align}
    r^G_t =& \ 0.1 r^\mathrm{cv}_t + 0.1 r^\mathrm{cp}_t + 0.3 r^\mathrm{bv}_t + 0.5 r^\mathrm{bp}_t \\
    r^\mathrm{cv}_t =& \ \mathrm{exp}\left(-1.5 \ \mathrm{max}\left(0, \quad v^* - \rvd^\mathrm{ball}_t \cdot \dot{\rvx}^\mathrm{com}_t \right)^2 \right) \\
    r^\mathrm{cp}_t =& \ \mathrm{exp}\left(-0.5 \ ||\rvx^\mathrm{ball}_t - \rvx^\mathrm{com}_t||^2\right) \\
    r^\mathrm{bv}_t =& \ \mathrm{exp}\left(-\mathrm{max}\left(0, \quad v^* - \rvd^*_t \cdot \dot{\rvx}^\mathrm{ball}_t \right)^2 \right) \\
    r^\mathrm{bp}_t =& \ \mathrm{exp}\left(-0.5 \ ||\rvx^*_t - \rvx^\mathrm{com}_t||^2\right) .
\label{eqn:rewardDribbling}
\end{align}
$r^\mathrm{cv}_t$ and $r^\mathrm{cp}_t$ encourages the character to move towards and stay near the ball, where $\rvx^\mathrm{ball}_t$ and $\dot{\rvx}^\mathrm{ball}_t$ represent the position and velocity of the ball, $\rvd^\mathrm{ball}_t$ is a unit vector pointing from the character to the ball, and $v^* = 1\mathrm{m/s}$ is the target velocity at which the character should move towards the ball. Similarly, $r^\mathrm{bv}_t$ and $r^\mathrm{bp}_t$ encourages the character to move the ball to the target location, with $\rvd^*_t$ denoting a unit vector pointing from the ball to the target. The goal $\rvg_t = \tilde{\rvx}^*_t$ records the relative position of the target location with respect to the character. The state $\rvs_t$ is augmented with additional features that describe the state of the ball, including the position $\tilde{\rvx}^\mathrm{ball}_t$, orientation $\tilde{\rvq}^\mathrm{ball}_t$, linear velocity $\tilde{\dot{\rvx}}^\mathrm{ball}_t$, and angular velocity $\tilde{\dot{\rvq}}^\mathrm{ball}_t$ of the ball in the character's local coordinate frame.

\paragraph{Strike:}
Finally, to further demonstrate our approach's ability to compose diverse behaviors, we consider a task where the character's objective is to strike a target using a designated end-effector (e.g. hands). The target may be located at various distances from the character. Therefore, the character must first move close to the target before striking it. These distinct phases of the task entail different optimal behaviors, and thus requires the policy to compose and transition between the appropriate skills. The goal $\rvg_t = (\tilde{\rvx}^*_t, h_t)$ records the location of the target $\tilde{\rvx}^*_t$ in the character's local coordinate frame, along with an indicator variable $h_t$ that specifies if the target has already been hit. The task-reward is partitioned into three phases:
\begin{equation}
    r^G_t = 
    \begin{cases}
        1, & \text{target has been hit} \\
        0.3 \ r^\mathrm{near}_t + 0.3, & ||\rvx^* - \rvx_t^\mathrm{root}|| < 1.375m \\
        0.3 \ r^\mathrm{far}_t, & \text{otherwise} \\
    \end{cases} .
\end{equation}
If the character is far from the target $\rvx^*$, $r^\mathrm{far}_t$ encourages the character to move to the target using a similar reward function as the Target Location task (Equation~\ref{eqn:rewardLocation}). Once the character is within a given distance of the target, $r^\mathrm{near}_t$ encourages the character to strike the target with a particular end-effector,
\begin{align}
    r^\mathrm{near}_t & = 0.2\ \mathrm{exp}\left(-2 ||\rvx^* - \rvx^\mathrm{eff}_t||^2 \right) \nonumber + 0.8 \ \mathrm{clip}\left(\frac{2}{3} \rvd^*_t \cdot \dot{\rvx}_t^\mathrm{eff}, \ 0, \ 1 \right) ,
\end{align}
where $\rvx^\mathrm{eff}_t$ and $\dot{\rvx}_t^\mathrm{eff}$ denote the position and velocity of the end-effector, and $\rvd^*_t$ is a unit vector pointing from the character's root to the target. After striking the target, the character receives a constant reward of $1$ for the remaining time steps.

\paragraph{Obstacles:}
Finally, we consider tasks that involve visual perception and interaction with more complex environments, where the character's goal is to traverse an obstacle filled environment, while maintaining a target speed. Policies are trained for two types of environments. 1) An environment containing a combination of obstacles including gaps, steps, and overhead obstacles that the character must duck under. 2) An environment containing stepping stones that requires more precise contact planning. Examples of the environment are available in Figure~\ref{fig:teaser} and \ref{fig:framesTasks}. The task-reward is the same as the one used for the Target Heading task (Equation~\ref{eqn:rewardHeading}), except the target heading is fixed along the direction of forward progress. In order for the policy to perceive the upcoming obstacles, the state is augmented with a 1D height-field of the upcoming terrain. The height-field records the height of the terrain at 100 sample locations, uniformly spanning 10m ahead of the character.

\section{AMP Hyperparameters}
\label{sec:suppAmpParams}
Hyperparameter settings used in the AMP experiments are available in Table~\ref{tab:suppAMPParams}. For single-clip imitation tasks, we found that a smaller discount factor $\gamma = 0.95$ allows the character to more closely imitate a given reference motion. A larger discount factor $\gamma = 0.99$ is used for experiments that include additional task objective, since these tasks may require longer horizon planning, such as \emph{Dribble} and \emph{Strike}.

\begin{table}[h!]
{ \centering  
\caption{AMP hyperparameters.}
\label{tab:suppAMPParams}
\begin{tabular}{|l|c|}
\hline
{\bf Parameter} & {\bf Value}  \\ \hline
    $w^G$ Task-Reward Weight &  $0.5$  \\ \hline
    $w^S$ Style-Reward Weight &  $0.5$  \\ \hline
    $w^\mathrm{gp}$ Gradient Penalty &  $10$  \\ \hline
    Samples Per Update Iteration &  $4096$  \\ \hline
    Batch Size &  $256$  \\ \hline
    $K$ Discriminator Batch Size &  $256$  \\ \hline
    $\pi$ Policy Stepsize (Single-Clip Imitation)&  $2 \times 10^{-6}$  \\ \hline
    $\pi$ Policy Stepsize (Tasks)&  $4 \times 10^{-6}$  \\ \hline
    $V$ Value Stepsize (Single-Clip Imitation)&  $10^{-4}$  \\ \hline
    $V$ Value Stepsize (Tasks)&  $2 \times 10^{-5}$  \\ \hline
    $D$ Discriminator Stepsize &  $10^{-5}$  \\ \hline
    $\mathcal{B}$ Discriminator Replay Buffer Size &  $10^{5}$  \\ \hline
    $\gamma$ Discount (Single-Clip Imitation)&  $0.95$  \\ \hline
    $\gamma$ Discount (Tasks)&  $0.99$  \\ \hline
    SGD Momentum &  $0.9$  \\ \hline
    GAE($\lambda$) &  $0.95$  \\ \hline
    TD($\lambda$) &  $0.95$  \\ \hline
    PPO Clip Threshold &  $0.02$  \\ \hline
\end{tabular}
}
\end{table}

\section{Latent Space Model}
\label{sec:suppLatentSpace}

The latent space model follows a similar architecture as \citet{MCPPeng19} and \citet{merel2019neural}. During pretraining, an encoder $q(\rvz_t | \rvg_t)$ maps a goal $\rvg_t$ to a distribution over latent variables $\rvz_t$. A latent encoding $\rvz_t \sim q(\rvz_t | \rvg_t)$ is then sampled from the encoder distribution and passed to the policy as an input $\pi(\rva_t | \rvs_t, \rvz_t)$. The latent distribution is modeled as a Gaussian distribution $q(\rvz_t | \rvg_t) = \mathcal{N}(\mu_q(\rvg_t), \Sigma_q(\rvg_t))$, with mean $\mu_q(\rvg_t)$ and diagonal covariance matrix $\Sigma_q(\rvg_t)$. The encoder is trained jointly with the policy using the following objective:
\begin{align}
    \mathop{\mathrm{arg \ max}}_{\pi, q} \qquad \expec_{p(\tau | \pi, q)} \left[ \sum_{t=0}^{T-1} \gamma^t r_t \right] + \lambda \expec_{p(\rvg_t)} \left[\mathrm{D}_\mathrm{KL} \ \left[q(\cdot | \rvg_t) || p_0 \right] \right].
\end{align}
$\tau = \{(\rvs_t, \rva_t, \rvg_t, r_t)_{t=0}^{T-1}, \rvs_T, \rvg_T \}$ represents the goal-augmented trajectory, where the goal $\rvg_t$ may vary at each time step, and
\begin{align}
p(\tau | \pi, q) = & p(\rvg_0) p(\rvs_t) \prod_{t=0}^{T-1} \bigg( (\rvg_{t+1} | \rvg_t) p(\rvs_{t+1} | \rvs_t, \rva_t) \bigg. \\
& \bigg. \int_{\rvz_t} \pi(\rva_t | \rvs_t, \rvz_t) q(\rvz_t | \rvg_t) d\rvz_t \bigg)
\end{align}
is the likelihood of a trajectory under a given policy $\pi$ and encoder $q$. Similar to a VAE, we include a KL-regularizer with respect to a variational prior $p_0(\rvz_t) = \mathcal{N}(0, I)$ and coefficient $\lambda$. The policy and encoder are trained end-to-end with PPO using the reparameterization trick \citep{Kingma2014}. Once trained, the latent space model can be transferred to downstream tasks by using $\pi(\rva_t | \rvs_t, \rvz_t)$ as a low-level controller, and then training a separate high-level controller $u(\rvz_t | \rvs_t, \rvg_t)$ that specifies latent encodings $\rvz_t$ for the low-level controller. The parameters of $\pi$ are fixed, and a new high-level controller $u$ is trained for each downstream task.

During pretrainig, the latent space model is trained using a motion imitation, where the objective is for the character to imitate a corpus of motion clips. A reference motion is selected randomly at the start of each episode, and a new reference motion is selected every 5-10s. The goal $\rvg_t=(\hat{q}_{t+1}, \hat{q}_{t+2})$ specifies target poses from the reference motion at the next two time steps.

The networks used for $\pi$ and $u$ follow a similar architecture as the networks used for the policies trained with AMP. The encoder $q$ is modeled by a network consisting of two hidden layers, with 512 and 256 hidden units, followed by a linear output layer for $\mu_q(\rvg_t)$ and $\Sigma_q(\rvg_t)$. The size of the latent encoding is set to 16D. Hyperparameter settings are available in Table~\ref{tab:suppLatentParams}.

\begin{table}[h!]
{ \centering  
\caption{Latent space model hyperparameters.}
\label{tab:suppLatentParams}
\begin{tabular}{|l|c|}
\hline
{\bf Parameter} & {\bf Value}  \\ \hline
    Latent Encoding Dimension &  16 \\ \hline
    $\lambda$ KL-Regularizer &  $10^{-4}$  \\ \hline
    Samples Per Update Iteration &  $4096$  \\ \hline
    Batch Size &  $256$  \\ \hline
    $\pi$ Policy Stepsize (Pre-Training)&  $2.5 \times 10^{-6}$  \\ \hline
    $u$ Policy Stepsize (Downstream Task)&  $10^{-4}$  \\ \hline
    $V$ Value Stepsize &  $10^{-3}$  \\ \hline
    $\gamma$ Discount (Pre-Training)&  $0.95$  \\ \hline
    $\gamma$ Discount (Downstream Task)&  $0.99$  \\ \hline
    SGD Momentum &  $0.9$  \\ \hline
    GAE($\lambda$) &  $0.95$  \\ \hline
    TD($\lambda$) &  $0.95$  \\ \hline
    PPO Clip Threshold &  $0.02$  \\ \hline
\end{tabular}
}
\end{table}

\section{Spatial Composition}
Our experiments have so far focused primarily on \emph{temporal} compositions of skills, where a character performs different skills at different points in time in order to fulfill particular task objectives, such as walking to a target and then punching it. In this section, we explore settings that require \emph{spatial} composition of multiple skills, where the task requires a character to perform different skills simultaneously. To evaluate AMP in this setting, we consider a compositional task where a character needs to walk along a target heading direction while also waving its hand at a target height. The motion prior is trained using a dataset consisting of both walking motions and waving motions, but none of the motion clips show examples of walking and waving at the same time. Therefore, the onus is on the policy to spatially compose these different classes of skills in order to fulfill the two disparate objectives simultaneously.

In this task, the character has two objectives: 1) a target heading objective for moving along a target direction $\rvd^*$ at a target speed $v^*$, 2) and a waving objective for raising its right hand to a target height $y^*$. The goal input for the policy is given by $\rvg_t=(\tilde{\rvd}^*_t, v^*, y^*)$, with $\tilde{\rvd}^*_t$ being the target direction in the character's local coordinate frame.
The composite reward is calculated according to: 
\begin{equation}
    r^G_t = 0.5r^\mathrm{heading}_t + 0.5r^\mathrm{waving}_t,
\label{eqn:rewardSpatialComposition}
\end{equation}
where $r^\mathrm{heading}_t$ the same as the reward used for the Target Heading task equation~\ref{eqn:rewardHeading}, and $r^\mathrm{wave}_t$ is specified according to:
\begin{equation}
   r^\mathrm{wave}_t = \mathrm{exp}\left(-16 \left(y^\mathrm{hand}_t - y^*  \right)^2 \right),
\label{eqn:rewardWave}
\end{equation}
where $y^\mathrm{hand}_t$ is the height of character's right hand. 

To evaluate AMP's ability to compose disparate skills spatially, we compare policies trained using both walking and waving motions, with policies trained with only walking motions or only waving motions. Table~\ref{tab:perfWave} compares the performance of the different policies with respect to the target heading and waving objectives. Although the motion prior was not trained with any reference motions that show both walking and waving at the same time, the policy was able to discover behaviors that combine these different skills, enabling the character to walk along different directions while also waving its hand at various heights. The policies trained with only walking motions tend to ignore the waving objective, and exhibit solely walking behaviors. Policies trained with only the waving motion are able to fulfill the waving objective, but learns a clumsy shuffling gait in order to follow the target heading direction. These results suggest that AMP does exhibit some capability for spatial composition different skills. However, the policies trained with both datasets can still exhibit some unnatural behaviors, particularly when the target height for the hand is high.

\begin{table}[h!]
{ \centering  
\caption{Performance of policies trained using different dataset on a spatial compositional task that combines following a target heading and waving the character's hand at a target height. The normalized task returns for each objective is averaged across 100 episodes for each model. The model trained with both walking and waving motions achieves relatively high rewards on both objectives, while the models trained with only one type of motions perform well only on one of the objectives.}
\label{tab:perfWave}
\begin{tabular}{|l|c|c|}
\hline
{\bf Dataset (Size)} & {\bf Heading Return} & {\bf Waving Return} \\ \hline
    Wave (51.7s) & $0.683 \pm 0.195 $& $0.949 \pm 0.144$  \\ \hline
    Walk (229.7s) & $0.945 \pm 0.192$ & $0.306 \pm 0.378$  \\ \hline
    Wave + Walk (281.4s) & $0.885 \pm 0.184$ & $0.891 \pm 0.202$  \\ \hline
\end{tabular}
}
\end{table}

\begin{figure*}[t]
	\centering
    \subfigure{\includegraphics[height=0.365\columnwidth]{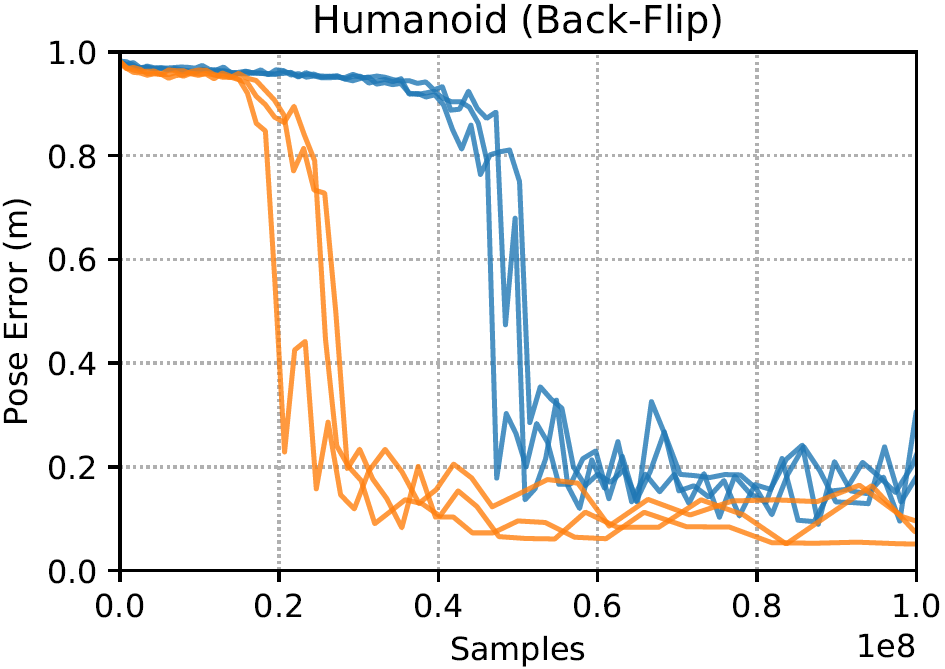}}
    \subfigure{\includegraphics[height=0.365\columnwidth]{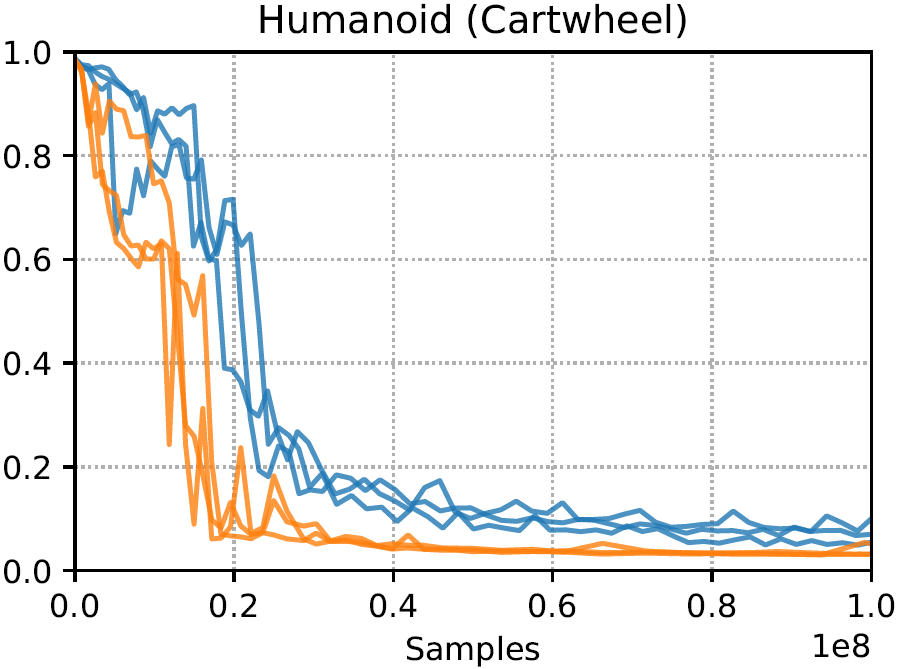}}
    \subfigure{\includegraphics[height=0.365\columnwidth]{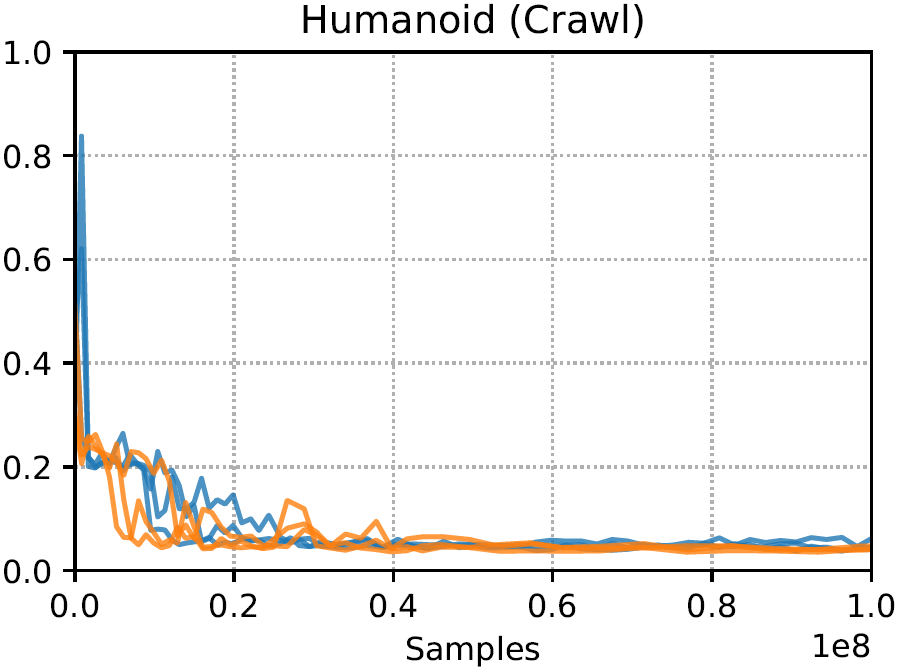}}
    \subfigure{\includegraphics[height=0.365\columnwidth]{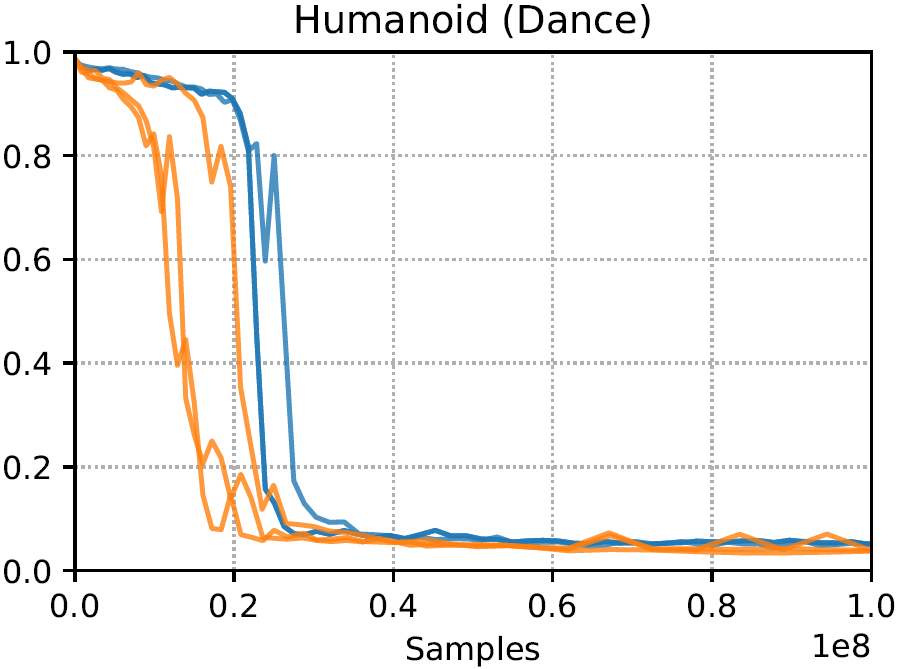}}\\
    \subfigure{\includegraphics[height=0.365\columnwidth]{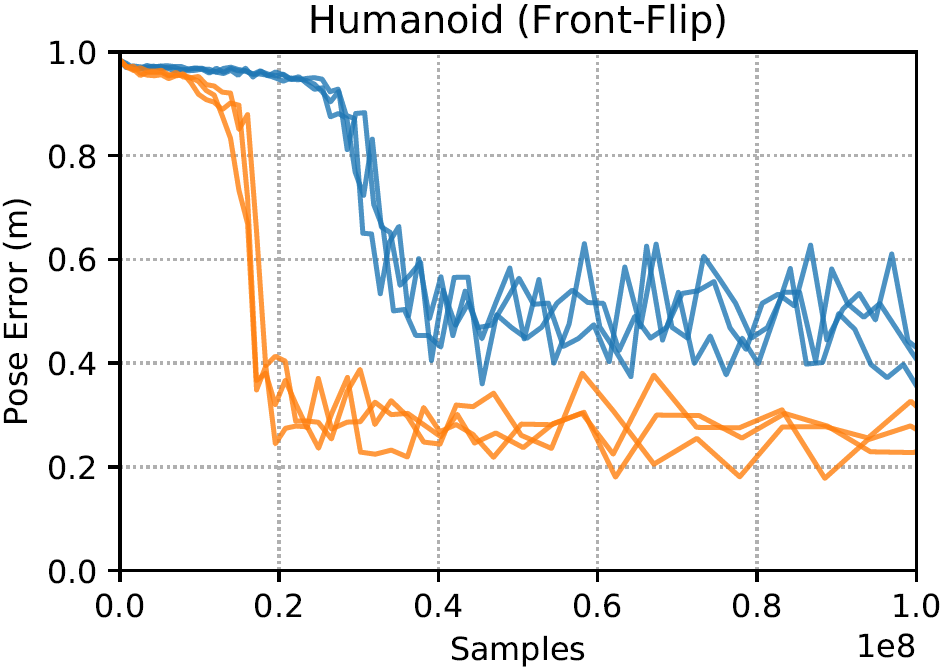}}
    \subfigure{\includegraphics[height=0.365\columnwidth]{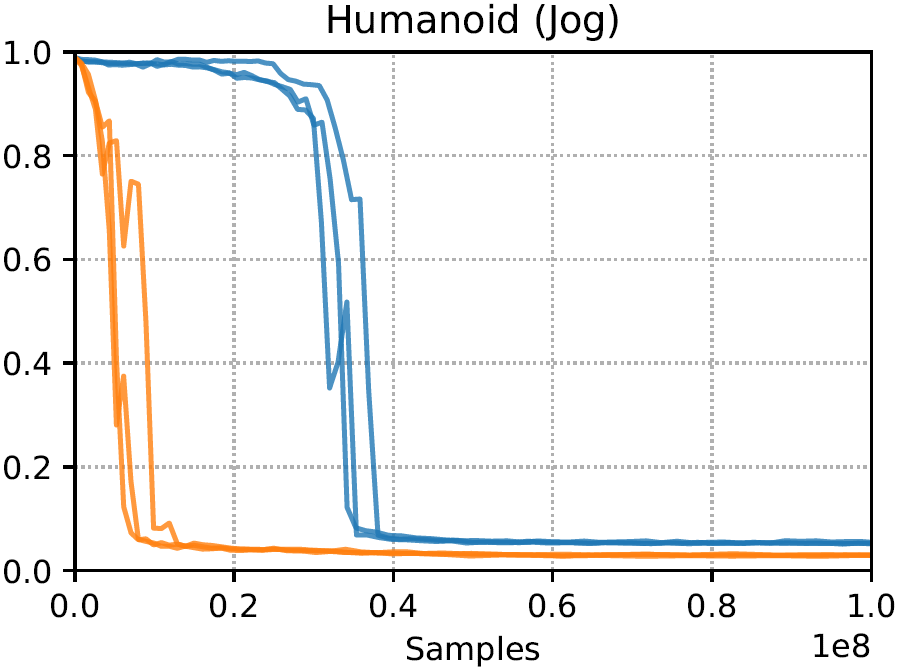}}
    \subfigure{\includegraphics[height=0.365\columnwidth]{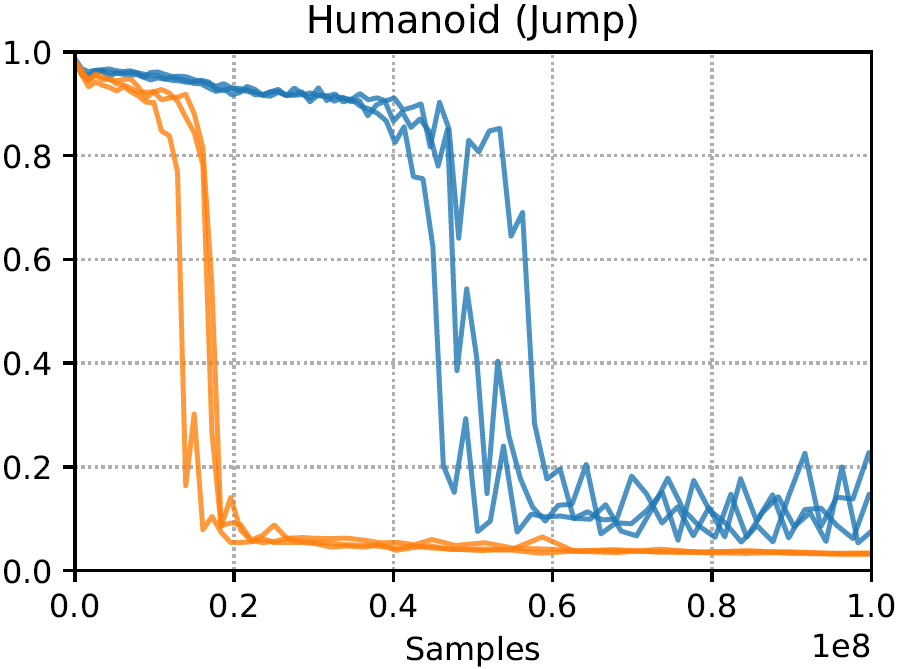}}
    \subfigure{\includegraphics[height=0.365\columnwidth]{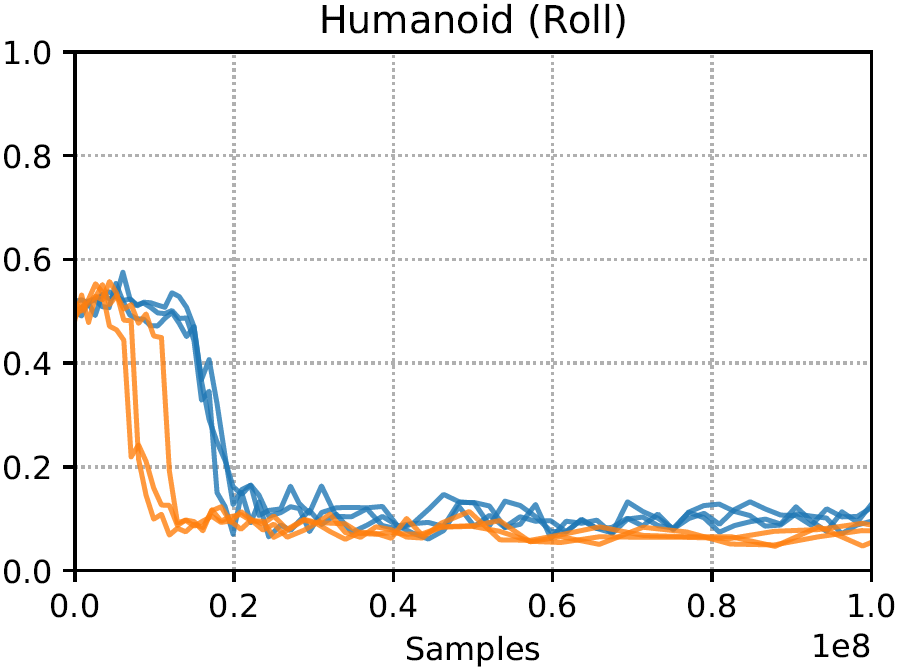}}\\
    \subfigure{\includegraphics[height=0.365\columnwidth]{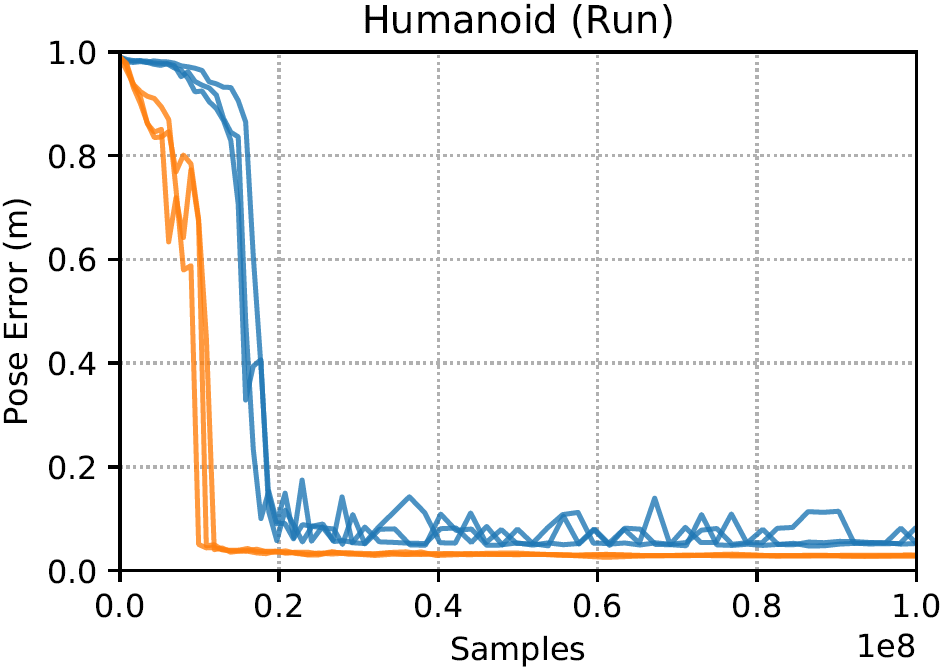}}
    \subfigure{\includegraphics[height=0.365\columnwidth]{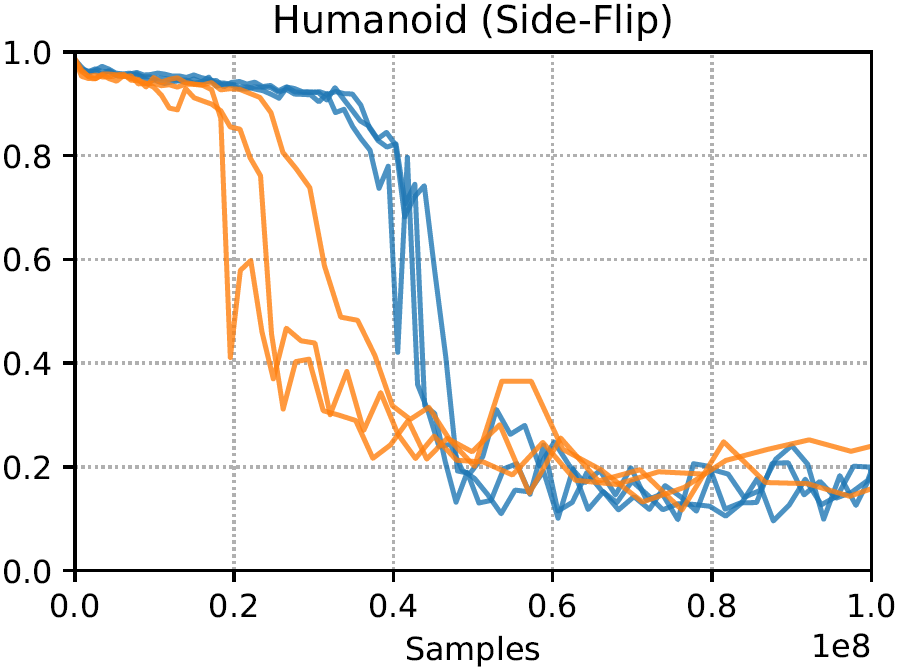}}
    \subfigure{\includegraphics[height=0.365\columnwidth]{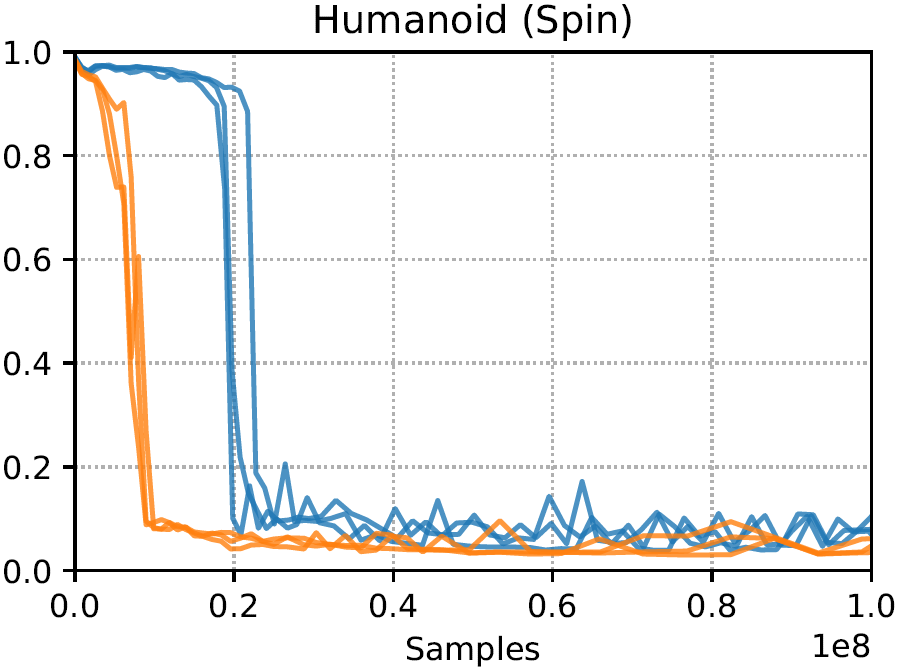}}
    \subfigure{\includegraphics[height=0.365\columnwidth]{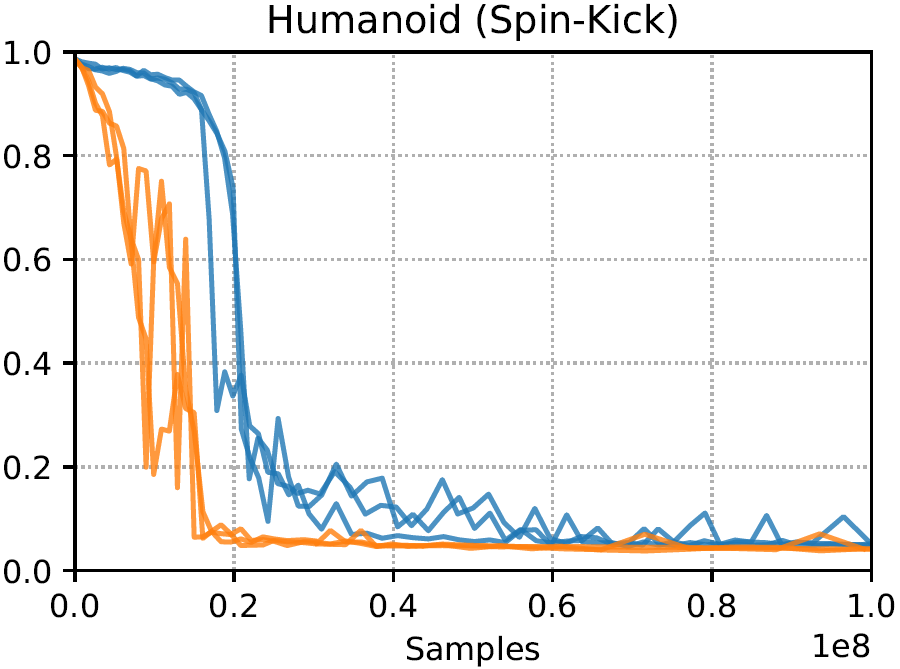}}\\
    \subfigure{\includegraphics[height=0.365\columnwidth]{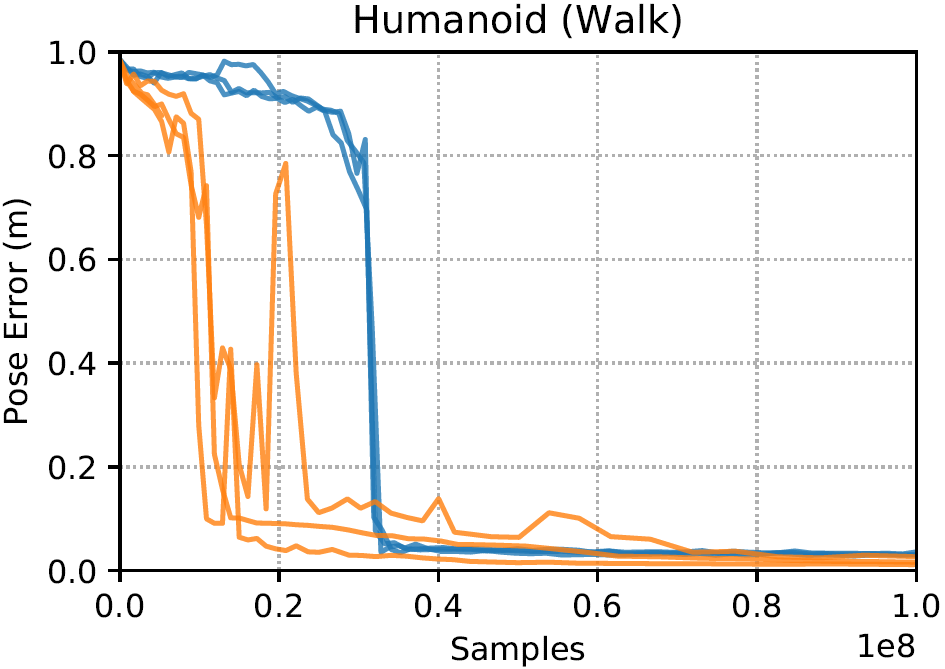}}
    \subfigure{\includegraphics[height=0.365\columnwidth]{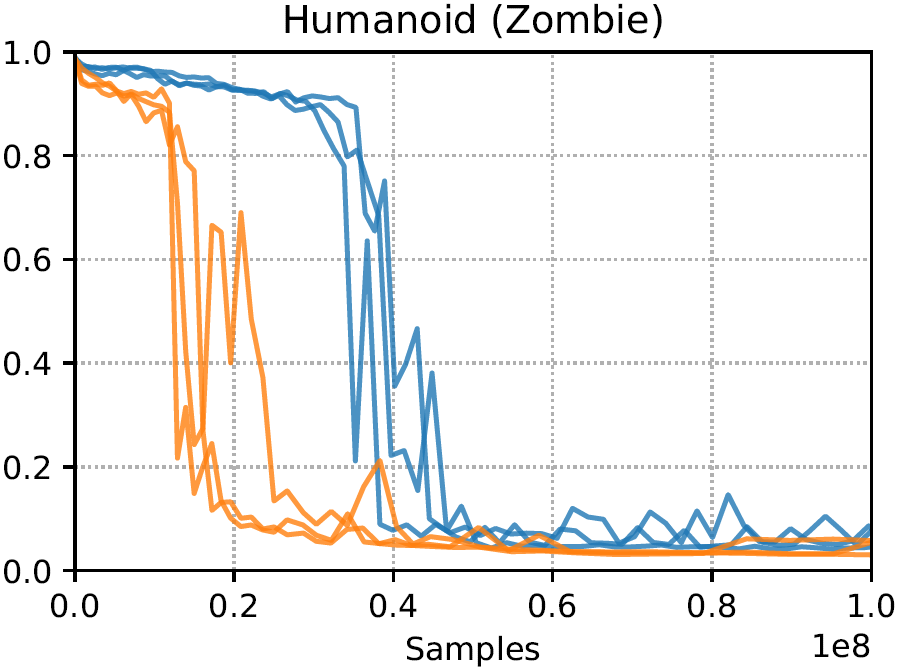}}
    \subfigure{\includegraphics[height=0.365\columnwidth]{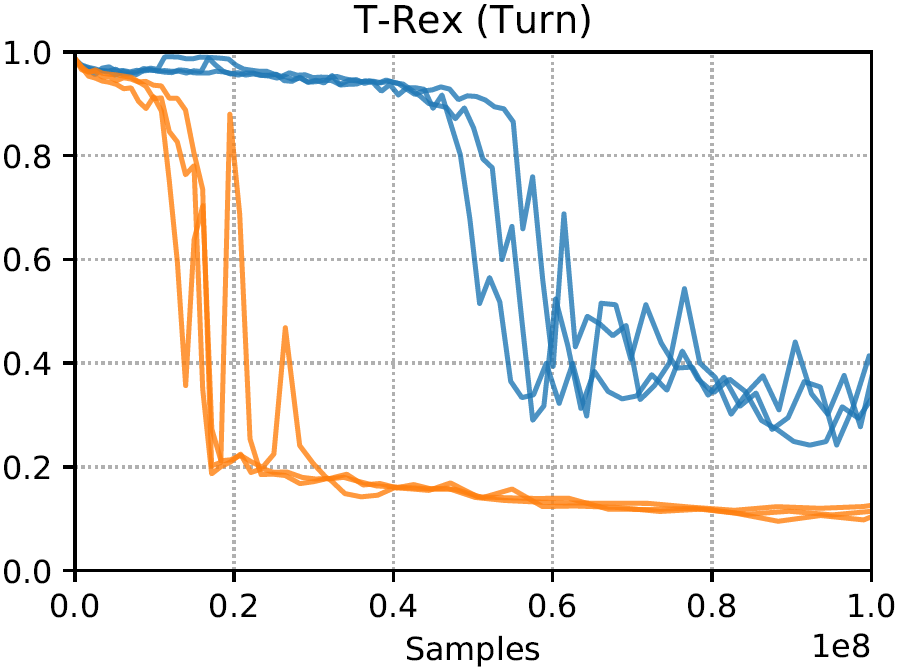}}
    \subfigure{\includegraphics[height=0.365\columnwidth]{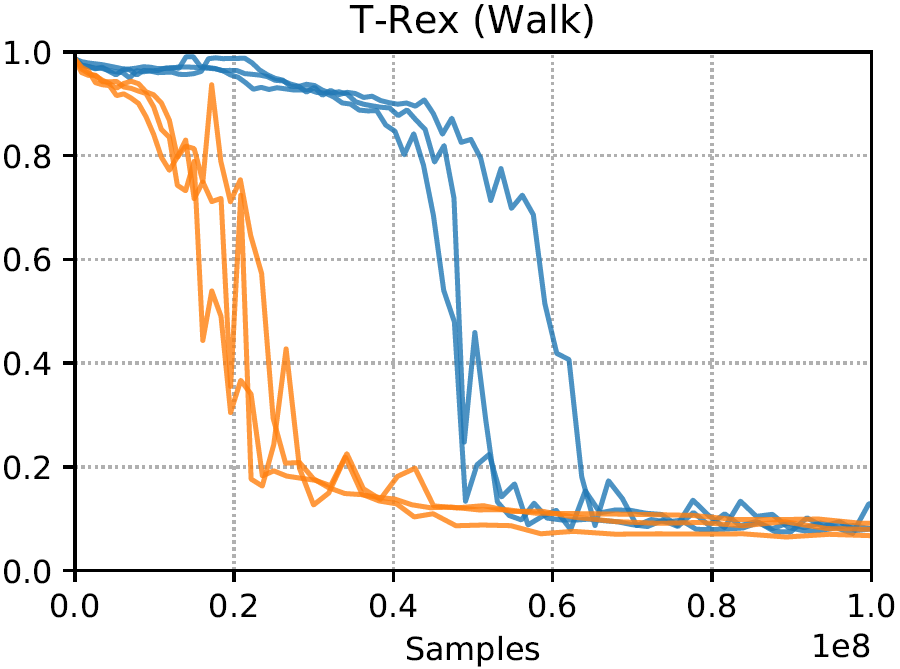}}\\
    \subfigure{\includegraphics[height=0.365\columnwidth]{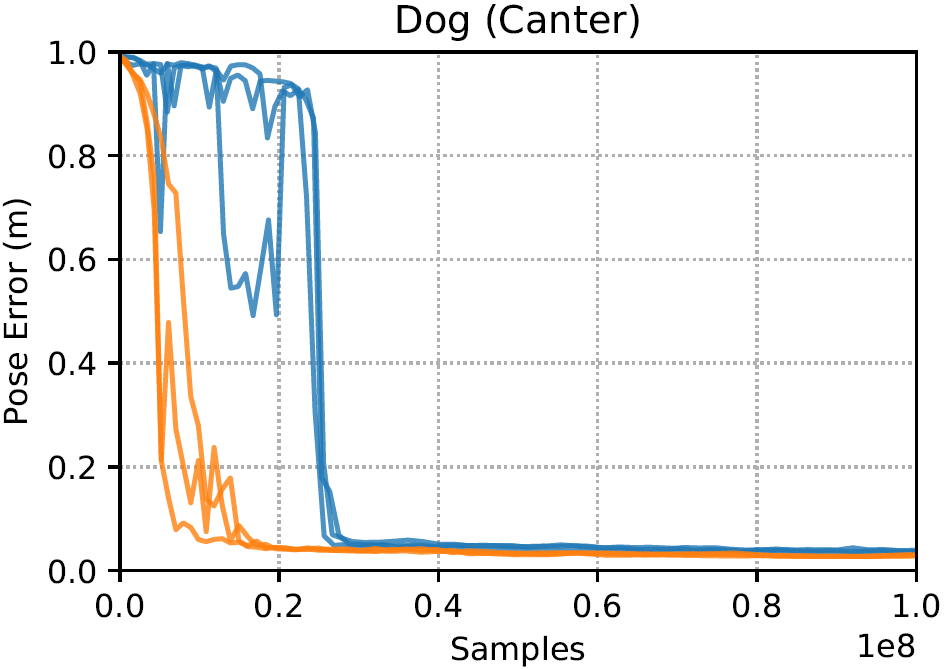}}
    \subfigure{\includegraphics[height=0.365\columnwidth]{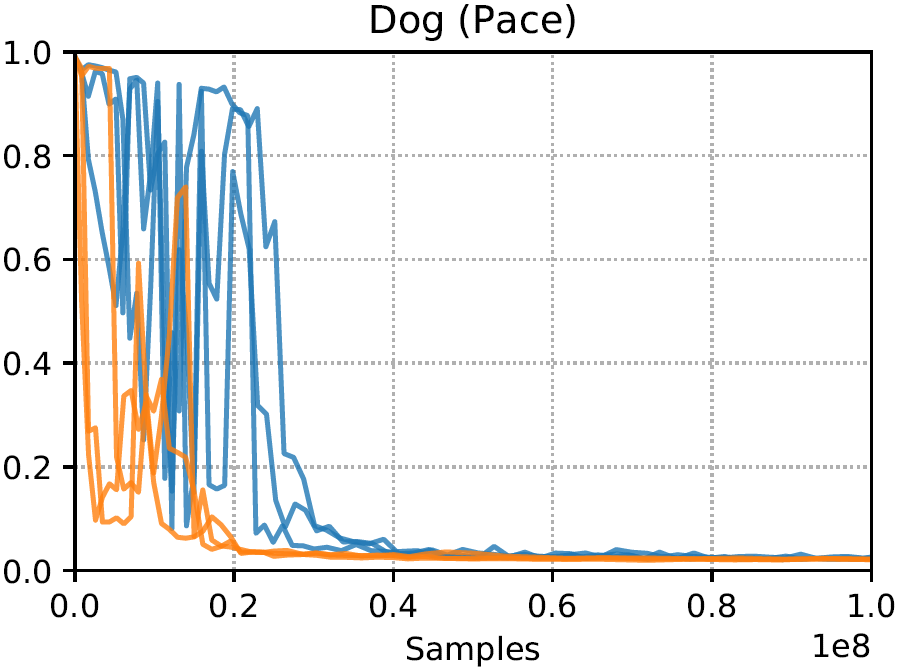}}
    \subfigure{\includegraphics[height=0.365\columnwidth]{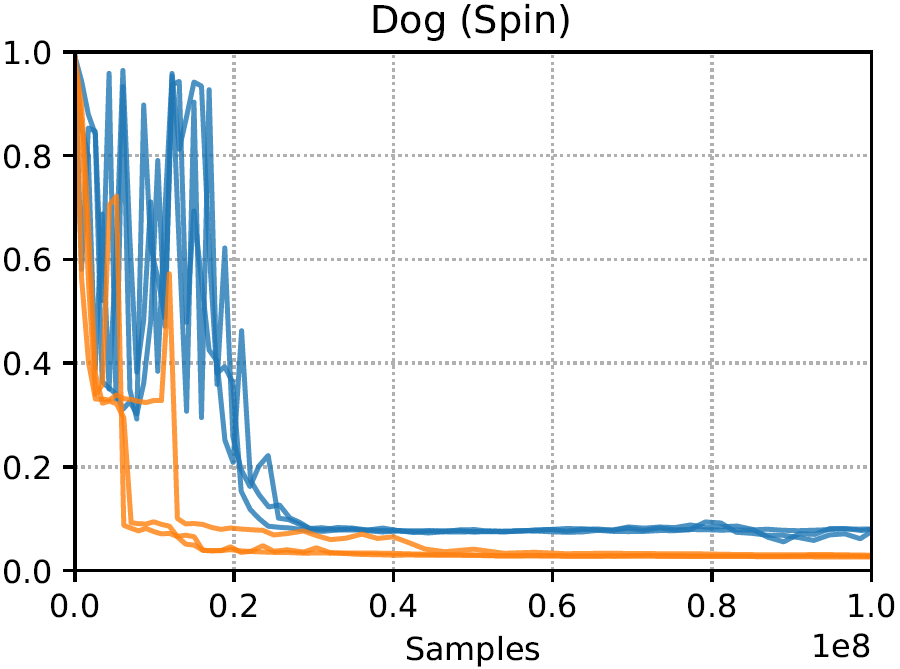}}
    \subfigure{\includegraphics[height=0.365\columnwidth]{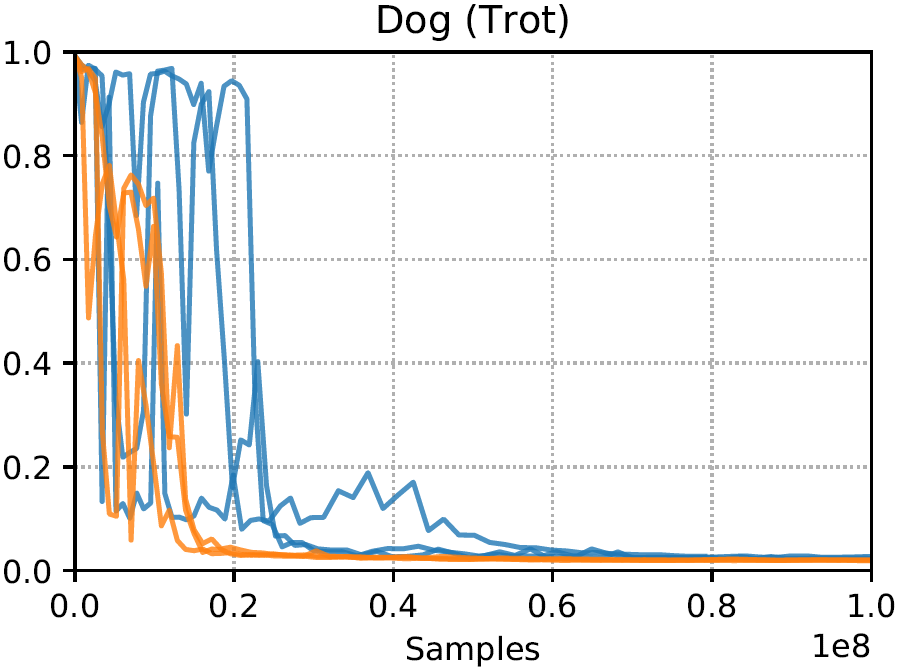}}\\
    \subfigure{\includegraphics[height=0.03\columnwidth]{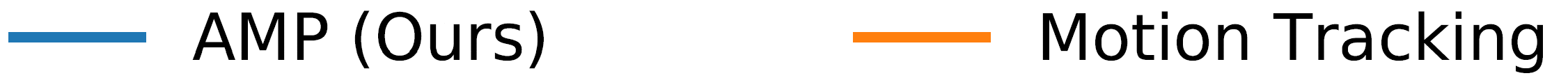}}\\
    \vspace{-0.25cm}
\caption{Learning curves comparing AMP to the motion tracking approach proposed by \citet{2018-TOG-deepMimic} (Motion Tracking) on the single-clip imitation tasks. 3 policies initialized with different random seeds are trained for each method and motion. AMP produces results of comparable quality when compared to prior tracking-based methods, without requiring a manually designed reward function or synchronization between the policy and reference motion.} 
\label{fig:suppCurvesImitation}
\end{figure*}

\begin{figure*}[t]
	\centering
    \subfigure{\includegraphics[height=0.365\columnwidth]{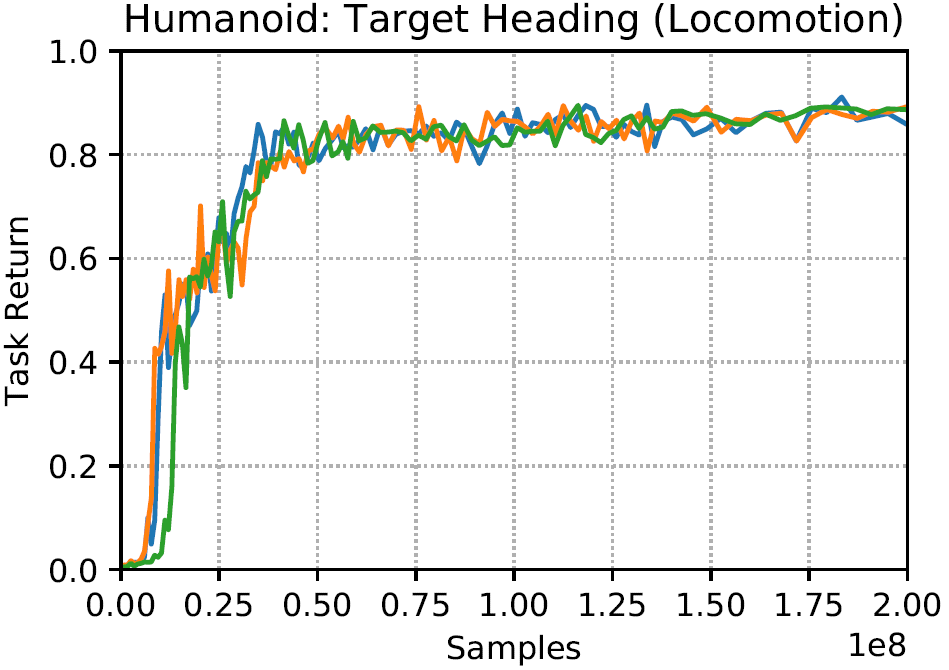}}
    \subfigure{\includegraphics[height=0.365\columnwidth]{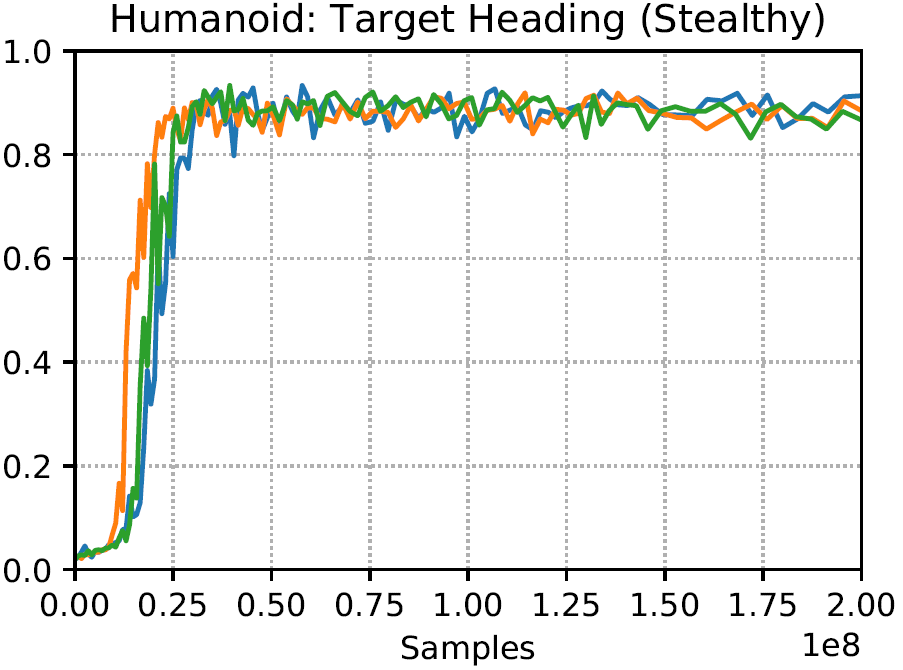}}
    \subfigure{\includegraphics[height=0.365\columnwidth]{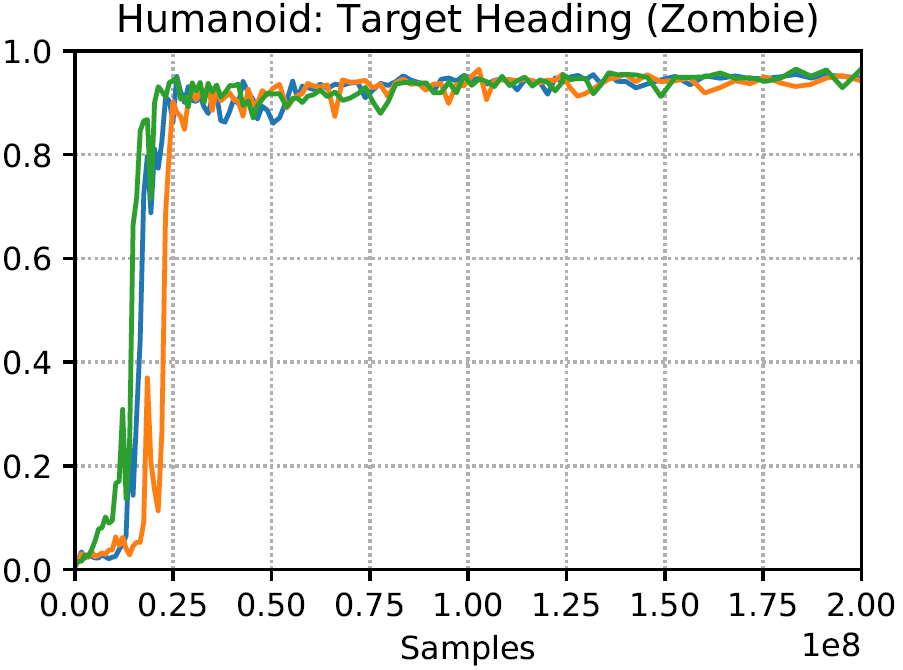}}
    \subfigure{\includegraphics[height=0.365\columnwidth]{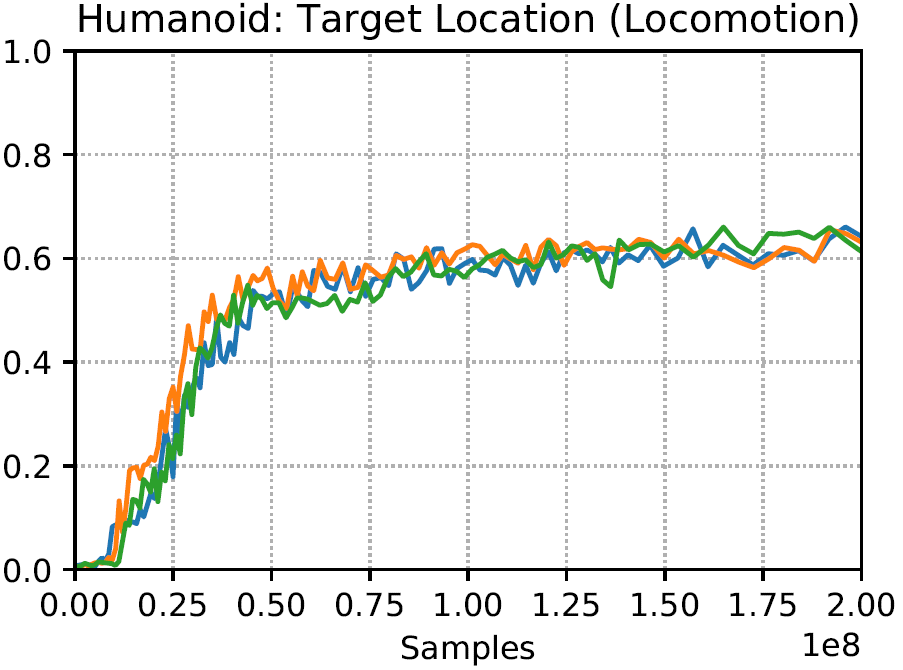}}\\
    \subfigure{\includegraphics[height=0.365\columnwidth]{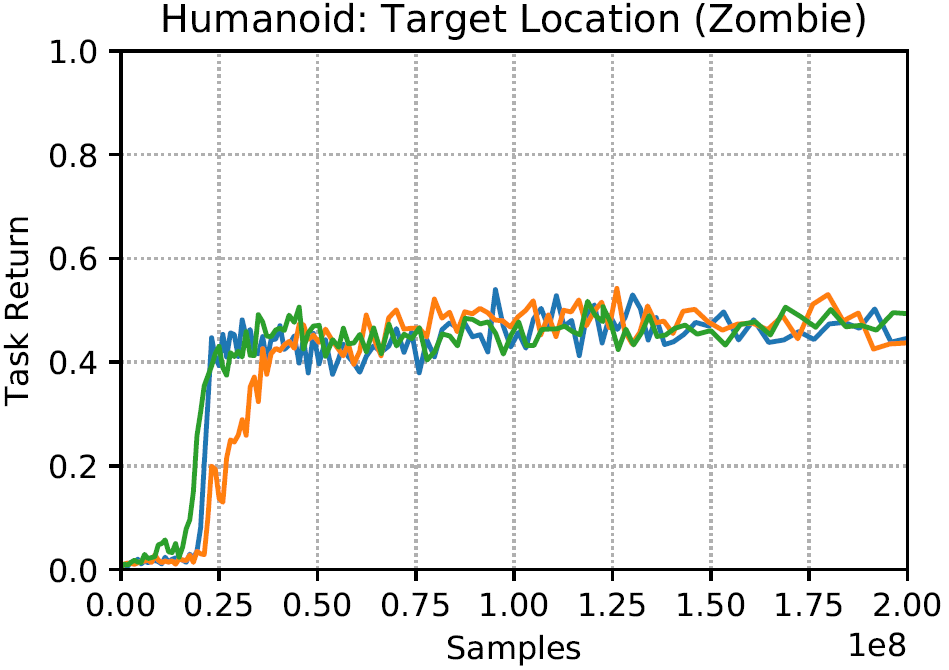}}
    \subfigure{\includegraphics[height=0.365\columnwidth]{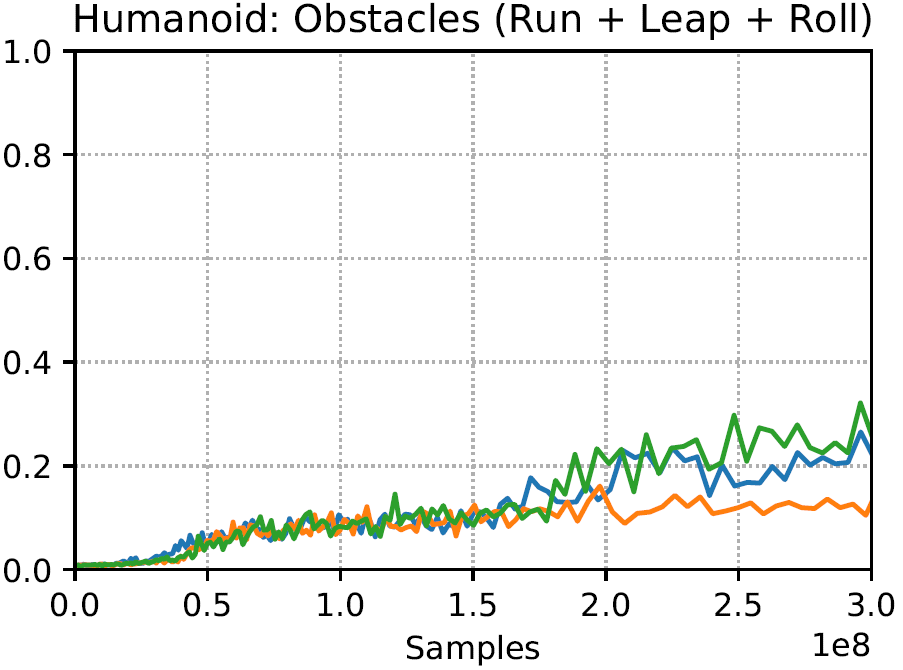}}
    \subfigure{\includegraphics[height=0.365\columnwidth]{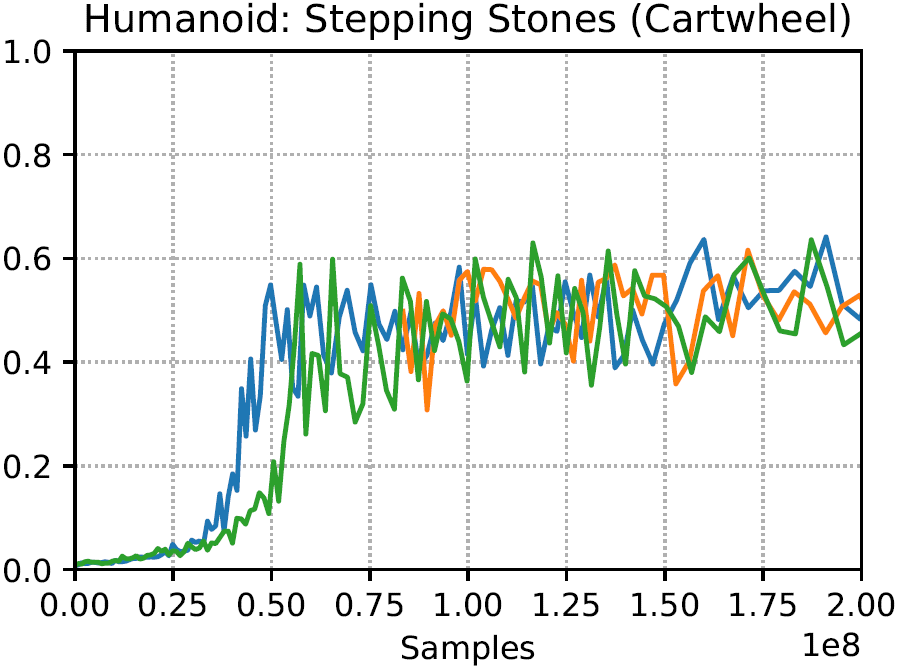}}
    \subfigure{\includegraphics[height=0.365\columnwidth]{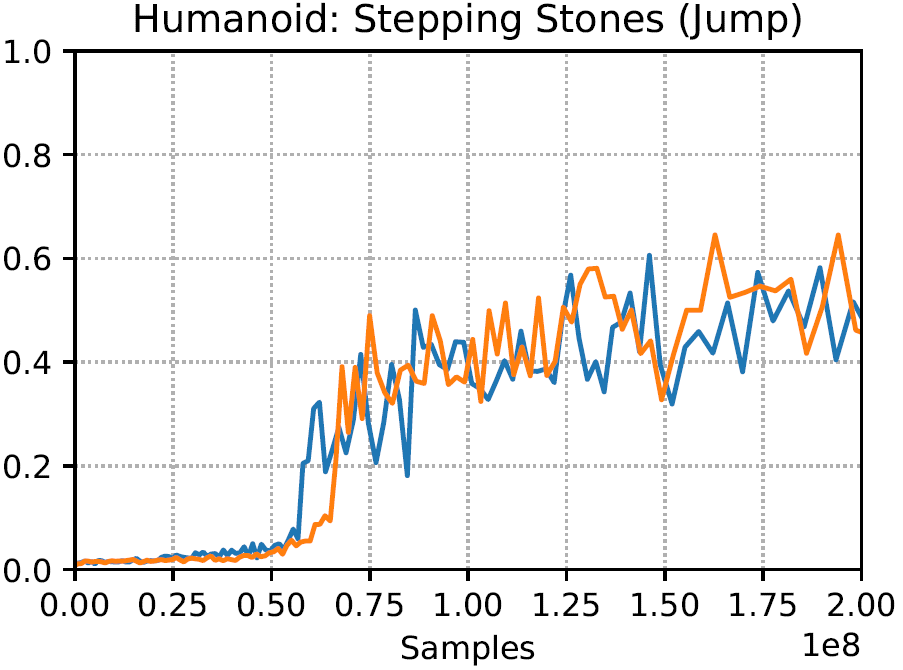}}\\
    \subfigure{\includegraphics[height=0.365\columnwidth]{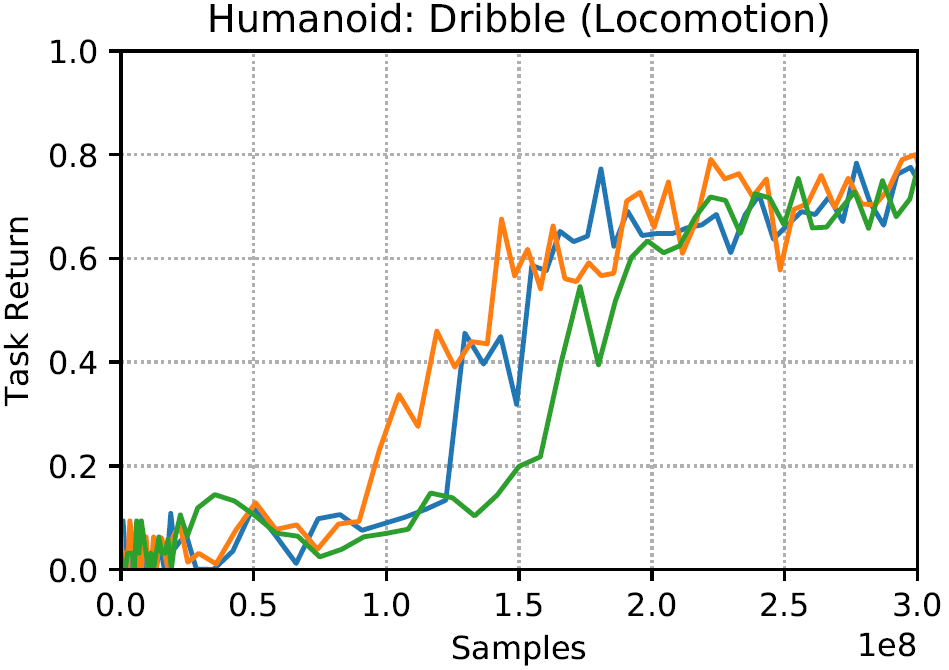}}
    \subfigure{\includegraphics[height=0.365\columnwidth]{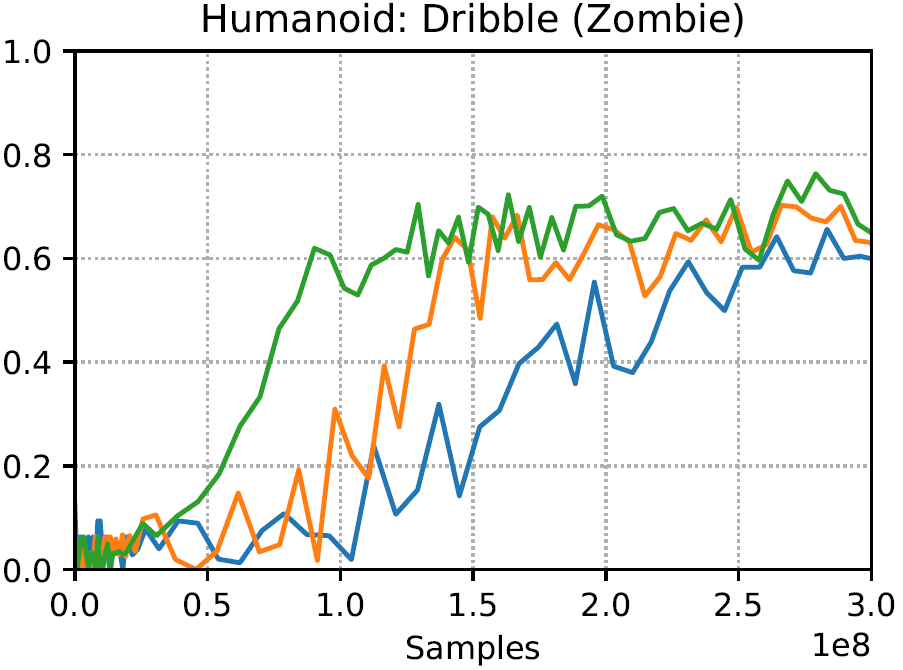}}
    \subfigure{\includegraphics[height=0.365\columnwidth]{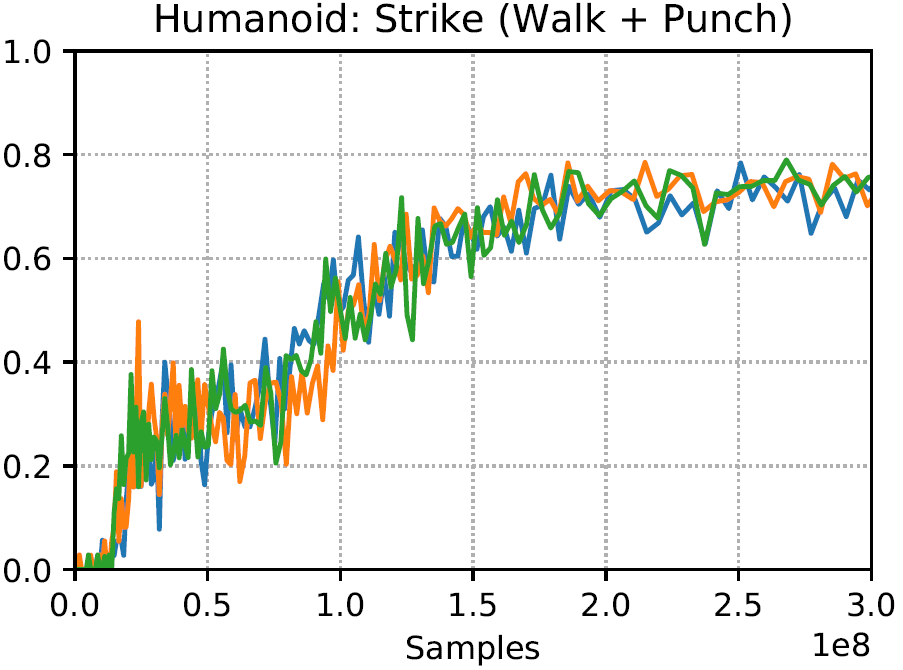}}
    \subfigure{\includegraphics[height=0.365\columnwidth]{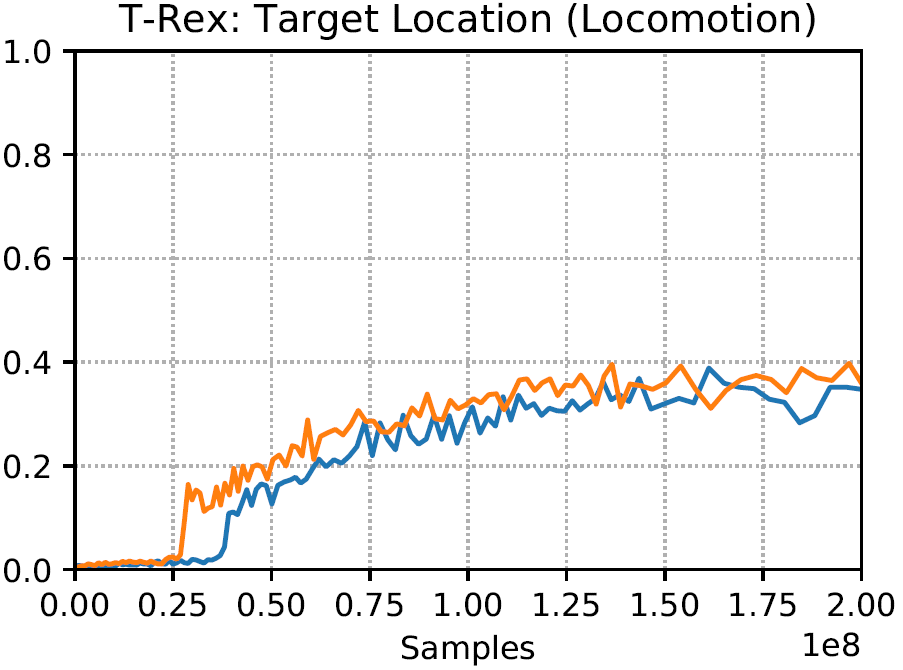}}\\
    \vspace{-0.25cm}
\caption{Learning curves of applying AMP to various tasks and datasets.} 
\label{fig:suppCurvesTasks}
\end{figure*}

\end{document}